%% file: main.tex
\documentclass[article, nojss]{jss}
\pdfoutput=1

\usepackage[utf8]{inputenc}
\PassOptionsToPackage{hyphens}{url}\usepackage{hyperref}
\hypersetup{
   breaklinks=true,   
   colorlinks=true,   
}
\usepackage{amsmath}
\usepackage{amssymb}
\usepackage[frozencache,cachedir=.]{minted}
\usepackage{minted}
\usepackage{amsthm}
\usepackage{array}
\usepackage{float}
\usepackage{pgfplots}
\pgfplotsset{compat=1.17}
\usepackage{rotating}
\usepackage[authoryear]{natbib}
\usepackage{booktabs} 
\usepackage{multirow, makecell}
\usepackage{dsfont}
\usepackage[toc,page]{appendix}
\usepackage{mathrsfs}
\usepackage{etoolbox}
\usepackage{algorithm}
\usepackage{algpseudocode}
\usepackage{layout}
\DeclareMathOperator*{\argmax}{arg\,max}
\DeclareMathOperator*{\argmin}{arg\,min}
\usepackage{mathtools}

\newcommand{\probP}{\text{I\kern-0.15em P}}

\newcommand{\approptoinn}[2]{\mathrel{\vcenter{
  \offinterlineskip\halign{\hfil$##$\cr
    #1\propto\cr\noalign{\kern2pt}#1\sim\cr\noalign{\kern-2pt}}}}}

\newcommand{\approxpropto}{\mathpalette\approptoinn\relax}

\newcommand{\specialcell}[2][c]{%
\renewcommand{\arraystretch}{1.0}
\begin{tabular}[#1]{@{}c@{}}#2\end{tabular}}

\newcommand*\samethanks[1][\value{footnote}]{\footnotemark[#1]}

\usepackage{framed}

\newcommand{\bl}{\\ & \qquad\qquad\qquad\qquad}

\Volume{VV}
\Year{YYYY}
\Month{MMMMMM}
\Issue{II}
\Submitdate{2023-08-25}
\Acceptdate{yyyy-mm-dd}

\author{Sacha Morin\thanks{Co-first authors} \\Université de Montréal 
\\ Mila - Quebec AI Institute 
   \And Robin Legault\samethanks \\Massachusetts Institute of Technology 
   \\ Operations Research Center
   \AND  Félix Laliberté\\  Université de Montréal
   \And Zsuzsa Bakk\\Leiden University
   \AND Charles-Édouard Giguère\\Institut universitaire en santé \\ mentale de Montréal
   \And Roxane \\\textbf{de la Sablonnière}\\Université de Montréal
   \And Éric Lacourse\\Université de Montréal
   }
\Plainauthor{Morin, Legault, Laliberté, Bakk, Giguère, de la Sablonnière \& Lacourse}
\title{\vspace{-0.45in}\pkg{StepMix}: A \proglang{Python} Package for Pseudo-Likelihood Estimation of Generalized Mixture Models with External Variables}
\Plaintitle{StepMix: A Python Package for Pseudo-Likelihood Estimation of Generalized Mixture Models with External Variables}
\Shorttitle{\pkg{StepMix}}

\Abstract{
    \pkg{StepMix} is an open-source \proglang{Python} package for the pseudo-likelihood estimation (one-, two- and three-step approaches) of generalized finite mixture models (latent profile and latent class analysis) with external variables (covariates and distal outcomes). In many applications in social sciences, the main objective is not only to cluster individuals into latent classes, but also to use these classes to develop more complex statistical models. These models generally divide into a measurement model that relates the latent classes to observed indicators, and a structural model that relates covariates and outcome variables to the latent classes. The measurement and structural models can be estimated jointly using the so-called one-step approach or sequentially using stepwise methods, which present significant advantages for practitioners regarding the interpretability of the estimated latent classes. In addition to the one-step approach, \pkg{StepMix} implements the most important stepwise estimation methods from the literature, including the bias-adjusted three-step methods with Bolk-Croon-Hagenaars and maximum likelihood corrections and the more recent two-step approach. These pseudo-likelihood estimators are presented in this paper under a unified framework as specific expectation-maximization subroutines. To facilitate and promote their adoption among the data science community, \pkg{StepMix} follows the object-oriented design of the \pkg{scikit-learn} library and provides an additional \proglang{R} wrapper.
}

\Plainkeywords{JSS, style guide, comma-separated, not capitalized, R}

\Address{
  Éric Lacourse\\
  Département de sociologie\\
  Faculté des arts et des sciences\\
  Université de Montréal\\
  C.P. 6128, succursale Centre-ville\\
  Montréal, QC\\
  H3C 3J7\\
  E-mail: \email{eric.lacourse@umontreal.ca}\\
}

\begin{document}

\section{Introduction}
\label{Sec:Introduction}
Mixture models are a family of probabilistic models that can be estimated from observed data to discover hidden or latent subgroups within a population. They are often used to analyze multivariate continuous and categorical data following some explicit assumptions about their conditional distribution given a categorical latent variable. Mixture modeling belongs to the field of model-based cluster analysis \citep{McLachlan2019} and takes different names across academic disciplines \citep{sterba2013understanding}. In health and social sciences, it is often referred to as latent class analysis when the observed variables are categorical and latent profile analysis when they are continuous \citep{oberski2016mixture}. Mixture models can also be presented under the more general framework of probabilistic graphical models as directed acyclic graphs with latent variables \citep{koller2009PGM}.

In social sciences, mixture models are used in cross-sectional and longitudinal studies that require to account for discrete population heterogeneity. For example, they are employed in psychology, psychiatry and epidemiology to establish probabilistic diagnoses when a gold standard is unavailable. Using national survey data, \cite{lacourse2010two} identified subgroups of adolescents that are more likely to present specific diagnoses of conduct disorder (non-aggressive vs. aggressive). Other applications based on cross-sectional data include identifying patterns of mobile internet usage for traveling \citep{okazaki15} and modeling recidivism rates among latent classes of juvenile offenders \citep{mulder2012}.

Historically, two main approaches have been used for estimating the parameters of mixture models. While it can be seen as a single likelihood maximization problem \citep{dayton}, the estimation process can also be divided into distinct steps. The first stepwise estimators of regression models became popular during the 1960s \citep{goldberger}. In order to relate latent classes to observed indicators and distal outcomes, these methods predict the class membership of the units based on the observed indicator variables before estimating the class-conditional distribution of the outcomes using the predicted latent classes. Their main advantage over classical maximum likelihood estimators (MLE) lies in the fact that they avoid the distal outcomes to contribute to the definition of the latent variables. This property is essential in applications where the interpretability of the latent classes is an important consideration. Among other advantages, stepwise methods are useful in the presence of missing data on distal outcomes, as the unobserved outcome variables only affect the last step of the estimation procedure. Unfortunately, several studies highlighted that naive stepwise methods produce highly biased estimates of the class-conditional distribution of the distal outcomes \citep{Croon_sem, bolck2004estimating, vermunt2010latent, Devlieger}. To alleviate this issue, stepwise estimators that mitigate or completely eliminate the bias introduced in the classification process have been proposed in the literature \citep{bolck2004estimating, vermunt2010latent,bakk2018two}.

Although commercial packages implement modern bias-adjusted stepwise estimators, these methods still have very limited availability in open-source software. Furthermore, they are completely unavailable in \proglang{Python} \citep{pythonlanguage}, which considerably reduces their pool of potential users. As the popularity of mixture modeling increases among applied researchers, so does the importance of making the state-of-the-art estimation methods publicly available in a well-documented software package. The goal of \pkg{StepMix} is to respond to this need by providing the scientific community with a convenient open-source package implementing bias-adjusted stepwise estimators. Our \proglang{Python} package is distributed via the \proglang{Python} Package Index (PyPI) at \url{https://pypi.org/project/stepmix/}. An \proglang{R} wrapper is also available on the Comprehensive \proglang{R} Archive Network (CRAN) at \url{https://cran.r-project.org/web/packages/stepmixr/index.html}.

This paper aims to introduce the \pkg{StepMix} package and present a self-contained reference on pseudo-likelihood estimation of mixture models with external variables. The family of mixture models studied in this work is formally defined in Section~\ref{Sec:Models}. Section~\ref{Sec:Related_packages} presents an overview of the estimation methods developed in the literature for these models and of the software packages in which they are implemented. Section~\ref{Sec:Pseudo_lik_estimation_methods} provides a detailed exposition of the estimation methods implemented in \pkg{StepMix} under a unified framework. The basic usage of our package and some of its most important features are discussed in Section~\ref{Sec:Using_StepMix}. Computational examples based on simulated and real-life datasets are presented in Section~\ref{Sec:Computational_examples}. Section~\ref{Sec:Conclusion} concludes the paper.

\section{Models}
\label{Sec:Models}

This section introduces the structure of the mixture models \pkg{StepMix} can estimate and the main notation used in the paper. In its most general formulation, the model we consider is composed of a latent categorical variable $X$, a set of observed indicators $Y$, a set of observed covariates (or predictors) $Z^p$, and a set of observed distal outcomes $Z^o$. The support of each of these sets of variables is respectively denoted by $\mathcal{X}$, $\mathcal{Y}$, $\mathcal{Z}^p$, and $\mathcal{Z}^o$. As illustrated in Figure~\ref{fig:CM}, the observed variables $Z^p$, $Y$, and $Z^o$ are assumed to be conditionally independent given the latent class $X$. 

\input{figures/fig_complete_model.tex}

Throughout the paper, for the sake of notational simplicity, the name of the variables are omitted in the formulas. For example, $p(x,y)$ denotes the joint probability $p(X{=}x, Y{=}y)$. Furthermore, the notation $p(\cdot)$ indistinctly refers to the probability mass function (PMF) in the case of discrete variables and to the probability density function (PDF) when handling continuous variables.

In accordance with the literature, we will distinguish the measurement model (MM) from the structural model (SM). They are respectively composed of the subsets of variables $(X,Y)$ and $(Z^p,X,Z^o)$. The MM is specified by a set of parameters $\theta_M \in \Theta_M$ that can be partitioned as $\theta_M = (\theta_X, \theta_Y) \in (\Theta_X \times \Theta_Y)$, where $\theta_X$ and $\theta_Y$ respectively specify the marginal distribution of the latent classes and the class-conditional distribution of the indicator variables.  The MM defines, for any realization $(x,y) \in \mathcal{X} \times \mathcal{Y}$, the joint probability:
\begin{equation}
   \label{eq:probasMM} p(x, y ; \theta_M) = p(x ; \theta_X)p(y | x ; \theta_Y)
\end{equation}

Analogously, we denote the parameters of the SM by $\theta_S = (\theta_{Z^p}, \theta_{Z^o})$, where the sets of parameters $\theta_{Z^p}$ and $\theta_{Z^o}$ respectively specify the distribution of the latent classes conditionally on the observed covariates and the class-conditional distribution of the distal outcomes. In the SM, the joint probability of a realization $(x,z^o) \in \mathcal{X} \times \mathcal{Z}^o$ given the covariates $z^p\in \mathcal{Z}^p$ is:
\begin{equation}
   \label{eq:probasSM_with_cov} p(x, z^o | z^p ; \theta_S) = p(x | z^p;\theta_{Z^p})p(z^o|x;\theta_{Z^o})
\end{equation}

In the absence of covariates, the latent class is treated as an exogenous variable in the SM. In this case, the SM's parameters reduce to $\theta_S = \theta_{Z^o}$ and only specify the class-conditional distribution of the distal outcomes.

Together, the MM and the SM form the complete model (CM), which specifies the probabilities:
\begin{equation}
   \label{eq:probasCM} p(x, y, z^o | z^p ; \theta_M,\theta_S) = p(x | z^p;\theta_X,\theta_{Z^p})p(y|x;\theta_Y)p(z^o|x;\theta_{Z^o})
\end{equation}

We adopt a generative perspective on the CM when it does not contain covariates. This means that the marginal distribution of all the variables, including the latent classes, is explicitly specified by the model. In particular, given fixed parameters $(\theta_X, \theta_Y, \theta_{Z^o})$, this allows to directly sample observations $(x,y,z^o)$ from the CM. On the other hand, a conditional perspective is adopted in the presence of covariates. In other words, the covariates $Z^p$ are regarded as exogenous variables, and their marginal distribution is not estimated. Observations $(x,y,z^o)$ can then be sampled from the CM given fixed covariates $z^p$ and parameters $(\theta_Y, \theta_{Z^p}, \theta_{Z^o})$. In this case, the parameters $\theta_X$ specifying the marginal distribution of the latent classes are ignored in the CM. The conditional distribution of the latent variable in Equation~\ref{eq:probasCM} thus reduces to:
\begin{equation}
   \label{eq:probasX}
   p(x | z^p ; \theta_X,\theta_{Z^p}) = 
   \begin{cases}
       p(x;\theta_X) & \text{if } Z^p = \emptyset,\\
       p(x | z^p ; \theta_{Z^p}) & \text{otherwise}.
   \end{cases}
\end{equation}

\section{Related packages}
\label{Sec:Related_packages}
Two classical approaches have been and are still widely applied in the literature for estimating the mixture models illustrated in Figure~\ref{fig:CM}. The first is the one-step approach, which directly estimates the parameters of the CM through likelihood maximization. This estimation is typically performed using the expectation-maximization (EM) algorithm \citep{dempster1977EM}. The second is the so-called naive three-step approach. Its first step consists of computing the MLE of the MM's parameters without considering the SM. In the second step, the latent class of each unit is predicted based on the estimated MM. In step three, after replacing the latent classes with their predicted values from step two, the parameters of the SM are estimated by likelihood maximization. The one-step and naive three-step approaches are based on standard EM, classification, and maximum likelihood estimation procedures. They can thus be implemented relatively easily based on any software for latent class modeling. Nevertheless, the commercial software packages \pkg{Latent Gold} \cite{vermunt2013technical} and \pkg{Mplus} \citep{muthen2017mplus} are the only ones that currently support three-step estimators natively.

The first bias-adjusted stepwise estimator that was introduced in the literature is the Bolk-Croon-Hagenaars three-step method (BCH), named after its authors \cite{bolck2004estimating}. The original BCH approach was improved by \cite{vermunt2010latent}, who also proposed the so-called ML three-step method. The main idea of bias-adjusted methods is to estimate the probability of misclassifying units in the second step in order to perform a correction of the imputed class weights used in step three. These methods are implemented in \pkg{Latent Gold} and \pkg{Mplus}, but are unavailable in open-source software.

An alternative stepwise estimator is the two-step approach \citep{bakk2018two}. This method bypasses the classification step of the three-step approach and thus eliminates the bias resulting from misclassification errors. It follows the logic of stepwise regression modeling \citep{goldberger} by first estimating the parameters of the MM before performing a pseudo-maximum likelihood estimation of the CM in which the estimated MM's parameters are held fixed. The two-step approach is available in \pkg{Latent Gold} and \pkg{Mplus}, and can easily be implemented in any software package allowing for conditioning on a set of fixed parameters. Unfortunately, most model-based clustering open-source packages do not allow using fixed parameters in this manner. The only exception is the recent \proglang{R} \citep{rlanguage} package \pkg{multilevLCA} \citep{multilevLCA}.

These stepwise estimation methods all fall within the frequentist paradigm of pseudo-maximum likelihood estimation. Although this is outside of the scope of this paper, it is worth mentioning that a Bayesian perspective on mixture modeling can also be adopted. In particular, maximum a posteriori (MAP) estimation makes it possible to include prior information in the analysis and is exploited as a regularization technique in model-based clustering \citep{fraley2007bayesian}. In addition to MAP estimation based on the EM algorithm, \pkg{BayesLCA} \citep{white2014bayeslca} supports Gibbs sampling \citep{geman1984stochastic} and a variational Bayes approximation method \citep{ormerod2010explaining}. The packages \pkg{depmixS4} \citep{visser2010depmixs4}, \pkg{randomLCA} \citep{beath2017randomlca}, \pkg{mclust} \citep{scrucca2016mclust} and \pkg{scikit-learn} \citep{pedregosa2011scikit} also implement Bayesian methods in the context of one-step estimation, primarily for regularization purposes.

A list of popular \proglang{R} and \proglang{Python} packages for the estimation of mixture models follows.
\begin{itemize}
    \item \pkg{mclust} \citep{scrucca2016mclust} is an \proglang{R} package implementing Gaussian mixture models with different covariance structures. The package also features functions for model-based hierarchical clustering and model selection. \pkg{AutoGMM} \citep{athey2019autogmm} is a \proglang{Python} package with similar features. 
    \item \pkg{MoEClust}~\citep{MurphyandMurphy2020} is an \proglang{R} package for parsimonious finite multivariate Gaussian mixtures of experts models. The package supports \pkg{mclust}’s different covariance structures and also gating/expert network covariates.
    \item \pkg{scikit-learn} \citep{pedregosa2011scikit} offers a number of Gaussian mixture models in \proglang{Python} and is a \pkg{StepMix} dependency.
    \item \pkg{multilevLCA} \citep{multilevLCA} is an \proglang{R} package for multilevel latent class models. It allows for the inclusion of covariates using the one-step and two-step estimators but does not support distal outcomes. It is currently the only other open-source package that implements a stepwise approach.
    \item \pkg{FlexMix} \citep{grun2008flexmix} is an \proglang{R} package that provides a flexible framework for finite mixture models of generalized linear models. It supports models with covariates as well as several families of probability distributions.
    \item \pkg{PoLCA} \citep{linzer2011polca} is an \proglang{R} package for one-step estimation of latent class models with categorical observed variables and covariates. It does not support Gaussian components.
    \item \pkg{MixMod} and \pkg{Rmixmod} \citep{lebret2015rmixmod} are the \proglang{Python} and \proglang{R} interfaces to the \proglang{C++} \pkg{mixmodLib} library. These packages are based on a generalized linear model perspective and implement model-based clustering methods for quantitative and qualitative data. They provide different variants of the EM algorithm, including the stochastic EM and the classification EM (see \cite{mclachlan2007algorithm} for a presentation of these methods). However, these packages do not support models with covariates.
\end{itemize}

In Table~\ref{tab:packagecomparison}, the main features of \pkg{StepMix} are compared with those of the previously mentioned software as well as the other open-source \proglang{R} and \proglang{Python} packages, \pkg{e1071::LCA}, \citep{meyer2019package}, \pkg{glca} \citep{kim2022glca}, and \pkg{VarSelLCM} \citep{marbac2019varsellcm}. Table~\ref{tab:packagecomparison} indicates the programming languages in which each package is available and, for \proglang{Python} packages, whether they follow a \pkg{scikit-learn} API. It also reports which packages natively support two-step and three-step estimation methods in addition to the standard one-step approach. The last two columns indicate whether each package supports both Gaussian and non-Gaussian indicators and outcome variables and whether the SM can contain covariates.

\input{tables/package_comparison}

\pkg{StepMix} is the first open-source software that natively implements bias-adjusted three-step methods. In addition to enabling the users of \proglang{Python} to use mixture models with covariates and both Gaussian and non-Gaussian components, \pkg{StepMix} is also the first package to support pseudo-likelihood estimation of mixture models in this language. One of our package's primary advantages over open-source and commercial existing software is that it strictly follows the \pkg{scikit-learn} interface. All the tools implemented in \pkg{scikit-learn}, including model selection and hyperparameter tuning methods, can thus be directly applied to \proglang{Python}'s \pkg{StepMix} estimator.

In social sciences, \pkg{Latent Gold} and \pkg{Mplus} are still the main tools used to estimate latent class models, mostly because of the lack of comprehensive packages in \proglang{Python} or \proglang{R} for mixture modeling. These software packages implement one-step, two-step, and bias-adjusted three-step estimators, with some limitations regarding the available distributions for non-categorical outcome variables. However, their code implementation of estimation methods is not open-source and thus cannot be directly inspected by researchers. Furthermore, these packages are expensive, making them inaccessible to a large part of the scientific community, hence the still frequent use of the one-step and naive three-step approaches in the literature. By bridging the gap between the features of commercial and open-source software \pkg{StepMix}, aims to foster the adoption of the state-of-the-art estimation methods of mixture models among applied researchers. 

\section{Pseudo-likelihood estimation methods}
\label{Sec:Pseudo_lik_estimation_methods}
The main objective of the estimation methods implemented in \pkg{StepMix} is to approximate the parameters of the generalized mixture model illustrated in Figure~\ref{fig:CM} using the maximum likelihood principle. In the estimation process, the MM and the SM can be considered either jointly or separately. This section presents the stepwise estimation methods from the literature under a unified framework in which each approach is based on the maximization of specific log-likelihood and pseudo-log-likelihood functions using the EM algorithm. The log-likelihood functions of the MM, SM, and CM are first presented in Section~\ref{Subsec:Likelihood}. Then, the EM algorithm is derived for the CM in Section~\ref{Subsec:EM}. In Sections~\ref{Subsec:One-Step} to \ref{Subsec:Three-Step}, the one-step, two-step, and three-step methods implemented in \pkg{StepMix} are described as specific EM subroutines that are applied on the MM, SM, and CM. This presentation highlights the fact that the stepwise estimation methods from the literature all share a very similar structure and primarily differ by the parameters that are considered fixed in the estimation process and, for three-step methods, by the definition of imputed class weights.

\subsection{Likelihood functions}
\label{Subsec:Likelihood}
Suppose we have a sample of units $i\in N$, each with observed covariates, indicators, and distal outcomes $(Z^p_i,Y_i,Z^o_i)=(z^p_i,y_i,z^o_i)$. Also, let the support of the latent variables $X_i$ be given by a set of $K$ classes $\mathcal{X}=\{1,\dots,K\}$. Finally, suppose that each unit has been given a sample weight $\omega_i \in \mathds{R}$. The log-likelihood function of the MM, SM, and CM are then respectively denoted by $\ell^{M}_N$, $\ell^{S}_N$ and $\ell^{C}_N$, and are given by:

\begin{align}
    \label{eq:ll_MM} \ell^{M}_N(\theta_M|y) &= \sum\limits_{i\in N} \omega_i\log p(y_i ; \theta_M)\\
    &= \sum\limits_{i\in N} \omega_i\log \left(\sum\limits_{k\in\mathcal{X}}p(X_i{=}k, y_i ; \theta_M)\right)\\
    &= \sum\limits_{i\in N} \omega_i\log \left(\sum\limits_{k\in\mathcal{X}}p(X_i{=}k; \theta_X)p(y_i | X_i{=}k ; \theta_Y) \right)
\end{align}
\begin{align}
    \label{eq:ll_SM} \ell^{S}_N(\theta_S| z^p, z^o ) &= \sum\limits_{i\in N} \omega_{i} \log p(z^o_i | z^p_i; \theta_S)\\
     &= \sum\limits_{i\in N} \omega_{i} \log \left(\sum\limits_{k\in\mathcal{X}}p(X_i{=}k,z^o_i| z^p_i; \theta_S)\right)\\
     &= \sum\limits_{i\in N} \omega_{i} \log \left(\sum\limits_{k\in\mathcal{X}}p(X_i{=}k | z^p_i ; \theta_{Z^p})p(z^o_i|X_i{=}k; \theta_{Z^o})\right)
\end{align}
\begin{align}
    \label{eq:ll_CM} \ell^{C}_N(\theta_M, \theta_S|y,z^o | z^p) &= \sum\limits_{i\in N} \omega_i \log p(y_i, z^o_i | z^p_i ; \theta_M, \theta_S)\\
    &= \sum\limits_{i\in N} \omega_i \log \left(\sum\limits_{k\in\mathcal{X}}p(X_i{=}k, y_i, z^o_i | z^p_i ; \theta_M, \theta_S)\right)\\
    \begin{split}
    &= \sum\limits_{i\in N} \omega_i \log \left(\sum\limits_{k\in\mathcal{X}}p(X_i{=}k | z^p_i ; \theta_X, \theta_{Z^p}) \right. \\
    & \qquad \qquad \left. \vphantom{\sum\limits_{k\in\mathcal{X}}} p(y_i | X_i{=}k ; \theta_Y)p(z^o_i|X_i{=}k; \theta_{Z^o})\right)
    \end{split}
\end{align}


\pkg{StepMix} supports missing data in the indicators and distal outcomes through full information maximum likelihood (FIML). This method was first outlined by \cite{hartley1971analysis}, and is a direct generalization of maximum likelihood estimation to datasets containing data points with missing observed variables. The FIML approach makes it possible to use all available information despite missing data by defining the probabilities in the likelihood function only with respect to the observed variables for each unit. Formally, let us consider a MM with $D$ independent indicators. Suppose that the indicators $y_{id}$ of unit $i \in N$ are missing for each dimension $d \in \bar{\mathcal{D}}_i \subseteq \mathcal{D} = \{1,\dots,D\}$. In the FIML framework, the probability $p(y_i | X_i{=}k ; \theta_Y) = \prod_{d \in \mathcal{D}} p(y_{id} | X_i{=}k ; \theta_Y)$ is replaced by $p(y_i | X_i{=}k ; \theta_Y) = \prod_{d \in \mathcal{D} \setminus \bar{\mathcal{D}}_i} p(y_{id} | X_i{=}k ; \theta_Y)$. If no indicator is observed, i.e., $\bar{\mathcal{D}}_i = \mathcal{D}$, then we fix $p(y_i | X_i{=}k ; \theta_Y) = 1$. In the absence of missing data, FIML reduces to classical likelihood maximization. Everywhere in the paper, including in the definition of the above log-likelihood functions, the probabilities $p(y_i | X_i{=}k ; \theta_Y)$ and $p(z^o_i|X_i{=}k; \theta_{Z^o})$ are therefore assumed to respect the definition required for the use of FIML.

\subsection{EM algorithm}
\label{Subsec:EM}
The log-likelihood functions of Section~\ref{Subsec:Likelihood} are generally multi-modal and non-convex due to the presence of a sum within the logarithm. The computation of their MLE is thus not straightforward and is tackled via the EM algorithm in \pkg{StepMix}.

The EM algorithm, introduced by \cite{dempster1977EM}, is the standard approach in the literature for maximum likelihood estimation in statistical models with latent categorical variables. The main idea of this method is to replace the log probability of each observation by a lower bounding auxiliary function to obtain a pseudo-likelihood function that is easier to maximize than the original log-likelihood function. 

In the following, we derive the EM algorithm for the CM. For each unit $i \in N$, the log probability of the observations $(y_i, z^o_i)$, given the covariates $z^p_i$ and the CM's parameters $(\theta_M, \theta_S)$, is underestimated by an auxiliary function $\mathcal{L}_i(q_i, \theta_M, \theta_S)$, which is parameterized by a probability distribution $q_i(\cdot)$.
\begingroup
\allowdisplaybreaks
\begin{align}
    \log p(y_i, z^o_i | z^p_i ; \theta_M, \theta_S) &= \log \left(\sum\limits_{x\in\mathcal{X}}p(x, y_i, z^o_i | z^p_i ; \theta_M, \theta_S)\right) \\
    \label{eq:function_q} &= \log \left(\sum\limits_{x\in\mathcal{X}} \frac{q_i(x) p(x, y_i, z^o_i | z^p_i ; \theta_M, \theta_S)}{q_i(x)} \right)\\
    &= \log \left(\mathds{E}_{q_i}\left[\frac{p(x, y_i, z^o_i | z^p_i ; \theta_M, \theta_S)}{q_i(x)}\right] \right)\\
    \label{ineq:Jensen} & \geq \mathds{E}_{q_i}\left[\log \left(\frac{p(x, y_i, z^o_i | z^p_i ; \theta_M, \theta_S)}{q_i(x)} \right)\right]\\
    &=: \mathcal{L}_i(q_i, \theta_M, \theta_S)
\end{align}
\endgroup

For Equation~\ref{eq:function_q} to hold, the only condition on $q(\cdot)$ is that  $q(x) > 0 \ \forall x \in \mathcal{X}$. Inequation~\ref{ineq:Jensen} results from Jensen's inequality, using the fact that the logarithm function is concave. 

From there, we want to select the probability distribution $q_i(\cdot)$ that maximizes the lower bounding function $\mathcal{L}_i(q_i, \theta_M, \theta_S)$ for it to approximate $\log p(y_i, z^o_i | z^p_i ; \theta_M, \theta_S)$ as closely as possible. To do so, we start by rewriting $\mathcal{L}_i(q_i, \theta_M, \theta_S)$ in terms of the Kullback–Leibler (KL) divergence of distribution $q(\cdot)$ from the conditional distribution of the latent variable $X_i$ given the observations of unit $i$ and the CM's parameters.
\begingroup
\allowdisplaybreaks
\begin{align}
 \mathcal{L}_i(q_i, \theta_M, \theta_S) &= \sum\limits_{x\in\mathcal{X}} q_i(x) \log \left(\frac{p(x, y_i, z^o_i | z^p_i ; \theta_M, \theta_S)}{q_i(x)} \right)\\
 &= \sum\limits_{x\in\mathcal{X}} q_i(x) \log \left(\frac{ p(x| z^p_i, y_i, z^o_i ; \theta_M, \theta_S) p(y_i, z^o_i | z^p_i ; \theta_M, \theta_S) }{q_i(x)} \right)\\
 \begin{split}
 &= \sum\limits_{x\in\mathcal{X}} q_i(x) \big( \log \left(\frac{p(x| y_i, z^o_i | z^p_i ; \theta_M, \theta_S)}{q_i(x)} \right) \bl +  \log\left( p(y_i, z^o_i  | z^p_i ; \theta_M, \theta_S) \right) \big)\\
 \end{split}\\
 \begin{split}
&= \sum\limits_{x\in\mathcal{X}} q_i(x) \log \left(\frac{p(x| z^p_i, y_i, z^o_i ; \theta_M, \theta_S)}{q_i(x)} \right)   
 \bl +  \sum\limits_{x\in\mathcal{X}} q_i(x)  \log\left( p(y_i, z^o_i  | z^p_i ; \theta_M, \theta_S) \right)  \\
\end{split}\\
\begin{split}
&= - \sum\limits_{x\in\mathcal{X}} q_i(x) \log \left(\frac{q_i(x)}{p(x| z^p_i, y_i, z^o_i ; \theta_M, \theta_S)} \right) \bl +  \sum\limits_{x\in\mathcal{X}} q_i(x)  \log\left( p(y_i, z^o_i  | z^p_i ; \theta_M, \theta_S) \right)  \\
\end{split}\\
\begin{split}
&= -D_{\text{KL}}(q_i(\cdot)||p(\cdot|z^p_i,y_i,z_i^o;\theta_M,\theta_S)) \bl  +  \sum\limits_{x\in\mathcal{X}} q_i(x)  \log\left( p(y_i, z^o_i | z^p_i ; \theta_M, \theta_S) \right)  \\
\end{split}\\
\begin{split}
&= -D_{\text{KL}}(q_i(\cdot)||p(\cdot|z^p_i,y_i,z_i^o;\theta_M,\theta_S)) \bl  +  \log\left( p(y_i, z^o_i  | z^p_i ; \theta_M, \theta_S) \right)\sum\limits_{x\in\mathcal{X}} q_i(x)  \\
\end{split}\\
\label{eq:aux_KL} &= -D_{\text{KL}}(q_i(\cdot)||p(\cdot|z^p_i,y_i,z_i^o;\theta_M,\theta_S))  +  \log\left( p(y_i, z^o_i | z^p_i ; \theta_M, \theta_S) \right)
\end{align}
\endgroup

Maximizing Equation~\ref{eq:aux_KL} with respect to $q(\cdot)$ over the set $\mathcal{Q}$ of probability distributions that are strictly positive over $\mathcal{X}$ is then relatively straightforward. 
\begingroup
\allowdisplaybreaks
\begin{align}
\begin{split}
\argmax_{q_i \in \mathcal{Q}} \left\{  \mathcal{L}_i(q_i, \theta_M, \theta_S) \right\} &= \argmax_{q_i \in \mathcal{Q}} \{  -D_{\text{KL}}(q_i(\cdot)||p(\cdot|z^p_i,y_i,z_i^o;\theta_M,\theta_S)) \bl  +  \log\left( p(y_i, z^o_i | z^p_i ; \theta_M, \theta_S) \right) \}\\
\end{split}\\
\label{eq:max_aux_q_1} &= \argmax_{q_i \in \mathcal{Q}} \left\{  -D_{\text{KL}}(q_i(\cdot)||p(\cdot|z^p_i,y_i,z_i^o;\theta_M,\theta_S)) \right\}\\
&= \argmin_{q_i \in \mathcal{Q}} \left\{  D_{\text{KL}}(q_i(\cdot)||p(\cdot|z^p_i,y_i,z_i^o;\theta_M,\theta_S) ) \right\}\\
\label{eq:max_aux_q_2} &= p(\cdot|z^p_i,y_i,z_i^o;\theta_M,\theta_S)
\end{align}
\endgroup
Equation~\ref{eq:max_aux_q_1} is obtained by removing the second term from the maximization problem. This can be done since this term does not depend on the distribution $q_i(\cdot)$ with respect to which the expression must be maximized. From there, the problem reduces to minimizing a KL divergence. We use the fact that the distribution that minimizes the KL divergence from a distribution $p(\cdot)$ is $p(\cdot)$ itself to obtain the result in Equation~\ref{eq:max_aux_q_2}.

After fixing the probability distributions $q_i(\cdot)$ for each unit $i \in N$, the log-likelihood function $\ell^{C}_N(\theta_M, \theta_S|z^p,y,z^o)$ of the CM can be approximated by replacing the log probability of each unit $i \in N$ by its auxiliary function $\mathcal{L}_i(q_i, \theta_M, \theta_S)$. The resulting pseudo-likelihood can then be maximized to approximate the MLEs of the original log-likelihood function.
\begingroup
\allowdisplaybreaks
\begin{align}
& \argmax_{(\theta_M, \theta_S) \in (\Theta_M, \Theta_S)} \left\{ \sum_{i \in N} \omega_i \mathcal{L}_i(q_i, \theta_M, \theta_S) \right\}\\ 
\begin{split}
&= \argmax_{(\theta_M, \theta_S) \in (\Theta_M, \Theta_S)} \left\{ \sum_{i \in N} \omega_i \left( \sum\limits_{x\in\mathcal{X}} q_i(x)  \log\left( p(x, y_i, z^o_i | z^p_i ; \theta_M, \theta_S) \right) \right. \right. 
\bl \left. \left. - \sum\limits_{x\in\mathcal{X}} q_i(x) \log \left( q_i(x) \right)  \right) \right\}\\
\end{split}\\
\begin{split}
&= \argmax_{(\theta_M, \theta_S) \in (\Theta_M, \Theta_S)} \left\{ \sum_{i \in N} \omega_i \left( \sum\limits_{x\in\mathcal{X}} q_i(x)  \log\left( p(x, y_i, z^o_i | z^p_i ; \theta_M, \theta_S) \right) \right)  
\right.  
\bl  \left.  
-  \sum_{i \in N} \omega_i \left(  \sum\limits_{x\in\mathcal{X}} q_i(x) \log \left( q_i(x) \right)  \right) \right\}\\
\end{split}\\
\label{eq:max_complete_information_theta} &= \argmax_{(\theta_M, \theta_S) \in (\Theta_M, \Theta_S)} \left\{ \sum_{i \in N} \omega_i \left( \sum\limits_{x\in\mathcal{X}} q_i(x)  \log\left( p(x, y_i, z^o_i | z^p_i ; \theta_M, \theta_S) \right) \right) \right\}
\end{align}
\endgroup

Analogously to Equation~\ref{eq:max_aux_q_1}, Equation~\ref{eq:max_complete_information_theta} is obtained by removing the terms that do not depend on the decision variables from the optimization problem. Equation~\ref{eq:max_complete_information_theta} corresponds to a likelihood maximization problem under complete information, as the latent variables $X_i$ are now treated as observed variables. The parameters $(\theta_M, \theta_S) \in (\Theta_M, \Theta_S)$ maximizing the objective function can thus be computed as in a standard MLE setting, without latent variables. 

The EM algorithm starts from an initial set of estimated parameters $(\hat{\theta}_M^{(0)}, \hat{\theta}_S^{(0)}) \in (\Theta_M \times \Theta_S)$. At each iteration, the distributions $q_i(x)$, $i \in N$ are updated using the incumbent estimates of the CM parameters. More precisely, for each unit $i \in N$ and each class $k \in \mathcal{X}$, we will denote the so-called class responsibilities by $\tau^{(t)}_{i_k} := q^{(t)}_i(X{=}k) = p(X{=}k|z^p_i,y_i,z_i^o; \hat{\theta}_M^{(t)}, \hat{\theta}_S^{(t)})$, which corresponds to the conditional distribution of $X_i$ given the observations of unit $i$ and the current estimated CM's parameters at iteration $t \in \{0,1,\dots\}$. This step is called the expectation step (E Step) since it corresponds to maximizing the expectation in Equation~\ref{ineq:Jensen}. The estimated CM's parameters are then updated by solving the complete information maximum likelihood problem given in Equation~\ref{eq:max_complete_information_theta} for the current imputed class weights $q^{(t)}_i(X{=}k) = \tau^{(t)}_{i_k}$. This latter step is referred to as the maximization step (M step).

This process is repeated iteratively until a user-defined convergence criterion is satisfied. As extensively discussed in \cite{Wu1983}, under mild conditions, the EM algorithm is guaranteed to converge to a local maximum of the original log-likelihood function. In practice, the EM algorithm is generally executed multiple times using different starting points to identify a global maximum with high probability. The convergence criteria implemented in \pkg{StepMix} are the relative and absolute gap between the average log-likelihood of subsequent estimated parameters. 

\newpage
\subsection{One-step method}
\label{Subsec:One-Step}
\input{pseudocode/1-step.tex}
For a set of units $i \in N$ with observed covariates, indicators, and distal outcomes $(Z^p_i, Y_i, Z^o_i)=(y_i,z^p_i,z^o_i)$, the objective of the one-step estimation method is to compute the MLE $(\hat{\theta}^C_M, \hat{\theta}^C_S) = \argmax_{(\theta_M, \theta_S) \in \Theta_M \times \Theta_S}\ell^{C}_N(\theta_M, \theta_S|z^p,y,z^o)$ of the CM's parameters. In \pkg{StepMix}, this is done by applying the EM algorithm on the CM directly, as presented in Algorithm~\ref{alg:1-step}. The notation $(\hat{\theta}^C_M, \hat{\theta}^C_S)$ makes explicit the fact that the parameters $\theta_M$ and $\theta_S$ of the MM and SM are estimated by maximizing the log-likelihood function $\ell^{C}_N(\cdot)$ of the CM in the one-step method. It intends to avoid notational ambiguity between this section and Sections~\ref{Subsec:Two-Step} and \ref{Subsec:Three-Step}. The lighter notation $(\hat{\theta}_M, \hat{\theta}_S)$ in used in the pseudocode.

The one-step approach is the most natural way to estimate the CM's parameters based on a single dataset in which the observable variables of both the MM and the SM are available for all the units. In this setting, the one-step approach is generally the method that minimizes the bias and the variance of the CM's parameter estimates. This has been illustrated in several computational studies \citep{vermunt2010latent, asparouhov14b, bakk2018two}, and is also visible in the results of Section~\ref{Sec:Computational_examples}. Nonetheless, as discussed in the same works, this method suffers from important practical and theoretical limitations. First, when using the one-step approach, model selection can be computationally cumbersome in the context of exploratory studies in which the number of potential indicator variables or covariates is large. Indeed, each time a variable is added to or removed from the model, both the SM and the MM must be reestimated. This can cause computation time issues when using complex models and large datasets. The main theoretical drawback of this method, however, lies in the fact that it makes the estimators of the MM, and thus the interpretation of the latent classes, dependent on the SM. As a consequence, researchers cannot properly study the relation between a given latent variable and multiple research questions using the one-step approach. Finally, this method requires both the MM and the SM's parameters to be estimated using the same set of units. In practice, this can lead to methodological issues and limit the quality of the estimators of the MM's parameters. For example, a group of researchers may publish a MM model estimated on a very large sample or census data without being allowed to share the data, for example, for privacy reasons. In this case, a second group of researchers wishing to use the same MM would need to reestimate it using their own dataset, possibly smaller, to estimate the parameters of their SM. The primary motivation of multi-step methods is thus to overcome these limitations by eliminating the dependency of the MM's parameters estimates on the SM.

\subsection{Two-step method}
\label{Subsec:Two-Step}
The two-step method requires a set of units $i \in N_1$ with observed indicators $Y_i=y_i$ and a set of units $j \in N_2$ with observed covariates, indicators, and distal outcomes $(Z^p_j, Y_j, Z^o_j)=(z^p_j,y_j,z^o_j)$. There are no further constraints on samples $N_1$ and $N_2$. In particular, $N_1$ and $N_2$ can be disjoint, but it is also possible that $N_1=N_2$ or $N_2 \subset N_1$. 

This approach aims at computing the MLE $\hat{\theta}^{M}_M = \argmax_{\theta_M \in \Theta_M}\ell^{M}_{N_1}(\theta_M|y)$ of the MM's parameters on sample $N_1$ and the maximum pseudo-likelihood estimator (MPLE) $\hat{\theta}^{C|M}_S = \argmax_{\theta_S \in \Theta_S}\ell^{C}_{N_2}(\theta_S|z^p,y,z^o ; \hat{\theta}_M)$ of the CM's parameters on sample $N_2$ conditionally on a fixed estimator $\hat{\theta}_M$ of the MM model parameters. 

In the first step, the EM algorithm is directly applied on the MM based on sample $N_1$. In the second step, the EM algorithm is applied on the CM for sample $N_2$, but the estimated MM's parameters $\hat{\theta}_M$ obtained in the first step are held fixed instead of being reestimated. This procedure is detailed in Algorithm~\ref{alg:2-step}.
\input{pseudocode/2-step.tex}

The idea of the two-step approach was first suggested by \cite{bandeen1997latent} as a method of evaluating the mutual sensitivity of the MM and SM's parameters in mixture models with discrete outcomes. It has then been applied by \cite{xue2002combining} in the context of SMs with covariates and by \cite{bartolucci2014comparison} in the case of latent Markov models for longitudinal data. Quite recently, \cite{bakk2018two} carried out the first comprehensive study of the two-step approach for generalized mixture models and provided empirical evidence for the superiority of the two-step approach over the widely used naive and bias-adjusted three-step methods in terms of bias and mean-squared error of the SM's parameters estimates. The two-step approach is an example of a two-stage pseudo-maximum likelihood estimation method \citep{besag1975statistical, gong1981pseudo}, and thus inherits from the theoretical properties of this class of estimators. In particular, this implies that the two-step estimators of the SM's parameters are consistent. For most applications in which the joint estimation of the MM and the SM's parameters is undesirable, the two-step approach should be preferred to three-step methods, as it avoids introducing a classification error that would negatively affect the quality of the SM's parameters estimates computed in the last step.

\subsection{Three-step method}
\label{Subsec:Three-Step}

Like the two-step method, the three-step approach is based on a set of units $i \in N_1$ with observed indicators $Y_i=y_i$ and a set of units $j \in N_2$ with observed covariates, indicators, and distal outcomes $(Z^p_j, Y_j, Z^o_j)=(z^p_j,y_j,z^o_j)$. Its first objective is also to compute the MLE $\hat{\theta}^{M}_M = \argmax_{\theta_M \in \Theta_M}\ell^{M}_{N_1}(\theta_M|y)$ of the MM's parameters based on sample $N_1$. From there, the goal is to compute a MPLE $\hat{\theta}^{S}_S = \argmax_{\theta_S \in \Theta_S}\tilde{\ell}^{S}_{N_2}(\theta_S|z;w)$ of the SM's parameters for sample $N_2$. The pseudo-log-likelihood function $\tilde{\ell}^{S}_{N_2}(\theta_S|z;w)$ is defined in a different way for each variant of the three-step method, but, in each case, only depends on the MM through a vector of imputed class weights $w_j \in \mathds{R}^K$ for each unit $j \in N_2$.

After applying the EM algorithm to the MM in the first step, imputed class weights are computed in the second step based on the estimated MM's parameters. In the naive version with soft assignment, the imputed class weights of each unit correspond to the class membership probabilities given by the MM. If modal assignment is used instead, each unit is completely assigned to its most likely latent class. The imputed class weights are then used in the third step to estimate the SM's parameters through a single iteration of the EM algorithm in which the responsibilities are fixed to the class weights obtained in the second step.

In the absence of uncertainty on the real latent classes, under usual regularity conditions on the distributions of the SM's variables, the naive three-step method would inherit the consistency and asymptotic normality properties of the MLE. However, the assignation phase of the three-step methods generally suffers from a non-negligible classification error rate which leads to biased estimates of the SM's parameters. In particular, \citet{bolck2004estimating} demonstrated that the uncorrected three-step approach underestimates the strength of the relationship between the distal outcomes variable and class membership. This can be intuitively understood by noticing that classification errors introduce units drawn from other classes in the subsample used to estimate the structural parameters of a given latent class. Thus, the higher the classification error rate, the more similar these subsamples become in terms of true class membership. As a consequence, the true conditional distribution of the outcomes given a class $c \in \mathcal{X}$ impacts the estimated conditional distribution of the outcomes for other classes $s \in \mathcal{X}$. In practice, high misclassification rates thus lead to excessively similar estimated class-conditional distributions.

Two bias-adjusted three-step methods, namely the BCH \citep{bolck2004estimating, vermunt2010latent} and ML \citep{vermunt2010latent, bakk2013} methods, have been developed to alleviate the downward bias of the naive three-step approach. They both include in their second step the computation of a left stochastic matrix $D \in \mathbb{R}^{K\times K}$ whose component $(c,k)$ contains the probability $p(W{=}k|X{=}c; \hat{\theta}_M)$ of assigning to class $k$ a unit that actually belongs to class $c$, given the estimated MM's parameters $\hat{\theta}_M$. As explained by \citet{vermunt2010latent}, if the support $\mathcal{Y}$ of the indicator variables $Y$ is finite, this probability can be computed exactly by summing over all the possible realizations of $Y$.
\begin{align*}
     D_{ck} = p(W{=}k|X{=}c; \hat{\theta}_M) = \frac{\sum_{y \in \mathcal{Y}} p(X{=}c|Y{=}y;\hat{\theta}_M)\bar{w}^{y}_{k}}{p(X{=}c; \hat{\theta}_M)}, \ \forall (c, k) \in \mathcal{X}^2
\end{align*}

Note that the definition of matrix $D$ depends on the type of assignment $\mathcal{A} \in \{\texttt{soft}, \texttt{modal}\}$ used in the second step through the weight $\bar{w}^{y}_{k}$ given to class $k$ for a unit with observed indicators $y$. This value is defined as follows.
\begin{align*}
\bar{w}^{y}_{k}  = \begin{cases}
p(X{=}k|Y{=}y;\hat{\theta}_M) & \text{, if $\mathcal{A}=\texttt{soft}$}, \\
\mathds{1}[\argmax_{c \in \{1,\dots,C\}}p(X|Y{=}y; \hat{\theta}_M) = k] & \text{, if $\mathcal{A}=\texttt{modal}$}.
\end{cases}
\end{align*}

In the presence of continuous indicators, or if $|\mathcal{Y}|$ is too large for the previous expression to be computed in a reasonable time, the probability $p(W{=}k|X{=}c; \hat{\theta}_M)$ can be estimated using the empirical distribution given by the observed sample $N_2$. This latter definition is natively used in \pkg{StepMix} for computational efficiency concerns. Matrix $D$ is thus computed as:
\begin{align*}
     D_{ck} = \hat{p}(W{=}k|X{=}c; \hat{\theta}_M) = \frac{\sum_{j \in N_2} p(X{=}c|Y{=}y_j;\hat{\theta}_M)w_{jk}}{p(X{=}c; \hat{\theta}_M)}, \ \forall (c, k) \in \mathcal{X}^2
\end{align*}
where $w_{jk} = \bar{w}^{y_j}_{k}$.

The third step of the BCH method is identical to that of the naive three-step method. It consists of estimating the SM’s parameters through a single iteration of the EM algorithm using imputed class weights. The only difference is that, in BCH, these weights are computed based on the inverse of matrix $D$ and, for each unit $j \in N_2$, are typically negative, except for the most probable class $k$ based on the observed indicators $y_j$ and the estimated MM's parameters $\hat{\theta}_M$. We refer the interested reader to \cite{vermunt2010latent, bakk2013} for a detailed discussion on the BCH method.

\input{figures/fig_structural_model_ML.tex}

In the case of the ML correction, the third step consists in splitting each unit $j \in N_2$ into $K$ copies $k \in \mathcal{X}$ with weight $w_{jk}$ and predicted class membership $W_j{=}k$, and applying the standard EM algorithm with multiple iterations on the model illustrated in Figure~\ref{fig:SM_ML}. This model is identical to the CM, except that the observed indicator variables are replaced by the predicted class membership, whose class-conditional distribution is given by $p(W{=}k|X{=}c; \hat{\theta}_M) = D_{ck}$.

At the $t^{\text{th}}$ iteration of the EM algorithm, the responsibility of the $k^{\text{th}}$ copy of unit $j \in N_2$, with predicted class membership $W_j{=}k$, is given by:
\begin{align}
    \tau^{(t)}_{j_kc} &= p(X_j{=}c|z^p_j,W_j{=}k,z_j^o; \hat{\theta}_M, \hat{\theta}_S^{(t)})\\
    &= \frac{p(W_j{=}k,z_j^o | z^p_j, X_j{=}c ; \hat{\theta}_M, \hat{\theta}_S^{(t)}) p(X_j{=}c | z^p_j ; \hat{\theta}_M, \hat{\theta}_S^{(t)})}{p(W_j{=}k,z_j^o | z^p_j; \hat{\theta}_M, \hat{\theta}_S^{(t)})} \\
    &= \frac{p(W_j{=}k| X_j{=}c ; \hat{\theta}_M ) p(z_j^o | z^p_j, X_j{=}c ; \hat{\theta}_S^{(t)}) p(X_j{=}c | z^p_j; \hat{\theta}_M, \hat{\theta}_S^{(t)})}{p(W_j{=}k,z_j^o | z^p_j ; \hat{\theta}_M, \hat{\theta}_S^{(t)})} \\
    &= \frac{D_{ck} p(z_j^o | z^p_j, X_j{=}c ; \hat{\theta}_S^{(t)}) p(X_j{=}c | z^p_j ; \hat{\theta}_M, \hat{\theta}_S^{(t)})}{p(W_j{=}k,z_j^o | z^p_j ; \hat{\theta}_M, \hat{\theta}_S^{(t)})} \\
    &= D_{ck} \frac{p(z_j^o | z^p_j, X_j{=}c ; \hat{\theta}_S^{(t)}) p(X_j{=}c | z^p_j ; \hat{\theta}_M, \hat{\theta}_S^{(t)})}{p(z_j^o | z^p_j ; \hat{\theta}_S^{(t)})} \frac{p(z_j^o | z^p_j ; \hat{\theta}_S^{(t)})}{p(W_j{=}k,z_j^o | z^p_j ; \hat{\theta}_M, \hat{\theta}_S^{(t)})}\\
    &= D_{ck} p(X_j{=}c | z^p_j, z_j^o ; \hat{\theta}_S^{(t)}) \frac{p(z_j^o | z^p_j ; \hat{\theta}_S^{(t)})}{p(W_j{=}k,z_j^o | z^p_j ; \hat{\theta}_M, \hat{\theta}_S^{(t)})}\\
    \label{eq:3_step_ML_prop} &\propto D_{ck} p(X_j{=}c | z^p_j, z_j^o ; \hat{\theta}_S^{(t)})p(z_j^o | z^p_j ; \hat{\theta}_S^{(t)})
\end{align}

In practice, to avoid having to handle a duplicated dataset in the third step of the three-step ML method, we express the responsibility $\tau^{(t)}_{jc}$ of class $c \in \mathcal{X}$ for each unit $j \in N_2$ as the weighted sum of the responsibilities $\tau^{(t)}_{j_kc}$.
\begin{align}
    \tau^{(t)}_{jc} &= \sum_{k \in \mathcal{X}} w_{jk}\tau^{(t)}_{j_kc}\\
    \label{eq:3_step_ML_exact} &= \sum_{k \in \mathcal{X}}  w_{jk} D_{ck} p(X_j{=}c | z^p_j, z_j^o ; \hat{\theta}_S^{(t)}) \frac{p(z_j^o | z^p_j ; \hat{\theta}_S^{(t)})}{p(W_j{=}k,z_j^o | z^p_j ; \hat{\theta}_M, \hat{\theta}_S^{(t)})} \\
    \label{eq:3_step_ML_approx} &\approxpropto \sum_{k \in \mathcal{X}}  w_{jk} D_{ck} p(X_j{=}c | z^p_j, z_j^o ; \hat{\theta}_S^{(t)})p(z_j^o | z^p_j ; \hat{\theta}_S^{(t)}) \\
    &= p(z_j^o | z^p_j ; \hat{\theta}_S^{(t)}) \sum_{k \in \mathcal{X}}  w_{jk} D_{ck} p(X_j{=}c | z^p_j, z_j^o ; \hat{\theta}_S^{(t)}) \\
    &\propto \sum_{k \in \mathcal{X}}  w_{jk} D_{ck} p(X_j{=}c | z^p_j, z_j^o ; \hat{\theta}_S^{(t)}) \\
    \label{eq:3_step_ML_final_resp} &= w_j D_c^T p(X_j{=}c | z^p_j, z_j^o ; \hat{\theta}_S^{(t)})
\end{align}

\input{pseudocode/3-step.tex}

Expression~\ref{eq:3_step_ML_approx} is obtained by replacing the class responsibilities $\tau^{(t)}_{j_kc}$ by their unnormalized form, given in Equation~\ref{eq:3_step_ML_prop} in the summation. This operation leads to an expression that is exactly proportional to Equation~\ref{eq:3_step_ML_exact} when using hard assignments, but not for soft assignments. However, in the latter case, it generally has a very small effect on the final class responsibilities. Those are obtained by normalizing Equation~\ref{eq:3_step_ML_final_resp}, as shown in Algorithm~\ref{alg:3-step}.

Although the three-step methods are generally dominated by the two-step approach regarding the bias and variance of the SM's parameters estimates \citep{bakk2018two}, they can present significant methodological advantages over the two-step approach for practitioners. Indeed, since the third step only depends on the MM through imputed class weights, estimating different SMs based on a three-step method does not require the researchers to have access to the indicator variables once the weights have been computed. For example, this allows the data owner to share only the relevant covariates, distal outcomes, and precomputed weight vectors with researchers interested in a particular structural model without disclosing the measurement model data. This property can be especially advantageous when the observed indicators used to estimate the measurement model contain sensitive data. 

Regarding the differences between the naive and bias-adjusted three-step methods, the sole advantage of the naive three-step approach is that its interpretation is very intuitive due to its explicit classification step. The BCH method is very similar in this regard, except that it produces imputed class weights that can be negative. Although slightly more complex to interpret and computationally more expensive than the other three-step approaches, the ML method generally produces the best estimators of the SM's parameters \citep{vermunt2010latent, bakk2013, bakk2018two}. Computational experiments carried out by \cite{bakk2013} indicate that soft assignments generally produce better estimates than modal assignments for all the three-step methods, although the difference in the resulting SM's parameters estimates is usually relatively small. 

\section[The StepMix package]{The \pkg{StepMix} package}
\label{Sec:Using_StepMix}
This section gives an overview of the \pkg{StepMix} package. Before detailing the supported estimators and providing code examples, we first discuss the \pkg{StepMix} documentation and how our package relies on the existing \proglang{Python} ecosystem.

\subsection{Software and documentation}
\pkg{StepMix} follows the object-oriented interface of the \pkg{scikit-learn} \proglang{Python} library \citep{pedregosa2011scikit}, therefore exposing an API that is both familiar to machine learning practitioners and intuitive for beginners and students. The strict adherence to the \pkg{scikit-learn} interface enables easy comparisons with other latent class/clustering methods in the \pkg{scikit-learn} ecosystem and grants \pkg{StepMix} users access to cross-validation iterators for model selection and hyperparameter tuning.

The \pkg{StepMix} source code was heavily inspired by, and depends on, the \pkg{scikit-learn} Gaussian mixture implementation. Most computations rely on vectorized operations over \pkg{NumPy} arrays \citep{harris2020array}, and \pkg{StepMix} estimators can be saved and loaded using standard \proglang{Python} tools, such as the \pkg{Pickle} module. Documentation follows the \pkg{NumPy}-style docstring convention and is compiled into a web page at \url{https://StepMix.readthedocs.io/en/latest/index.html}, which also features interactive notebook tutorials accessible to any user with a web browser.

\input{tables/distributions.tex}

\subsection{Estimators}
The simplest \pkg{StepMix} models do not include external variables (covariates and distal outcomes) and only have a single set of observed indicators. These models correspond to Gaussian or categorical mixtures, the latter of which is not currently available in \pkg{scikit-learn} (as of version 1.2). For example, assuming some integer-encoded categorical data is stored in the array \code{Y}, either as a \pkg{NumPy} array or a \pkg{Pandas} data frame \citep{mckinney-proc-scipy-2010, reback2020pandas}, a \pkg{StepMix} estimator for three latent classes can be declared and fit with the following commands.

\begin{minted}{python}
model = StepMix(n_components = 3, measurement = "categorical")
model.fit(Y)
\end{minted}

The most important parameters of a \pkg{StepMix} estimator are the number of latent classes (\code{n\_components)} and the specified conditional distribution (\code{measurement}). All options and distributions for the \code{measurement} argument are detailed in Table~\ref{tab:StepMix_models}. Additional parameters of interest include seeding (\code{random\_state}) and various EM optimization parameters, such as the maximum number of EM iterations (\code{max\_iter}), the tolerance for stopping the optimization (\code{abs\_tol}) and the number of different initializations to try (\code{n\_init}). Users can obtain a detailed model report presenting parameters and fit statistics (\code{verbose=1}), examples of which are provided in Appendix~\ref{Sec:Appendix}.

Adding a second set of observed variables and specifying a SM is a simple matter of adding a \code{structural} argument (the options of which are also detailed in Table~\ref{tab:StepMix_models}). For the SM, \pkg{StepMix} relies on the familiar \pkg{scikit-learn} supervised learning interface: the SM data is provided as a second set of variables, typically reserved for target labels. This allows a clear separation between the MM and the SM data, and makes a clear distinction in the case where only a MM is specified.

The specific stepwise estimation procedure (\code{n\_steps}) (see Section~\ref{Sec:Pseudo_lik_estimation_methods}) and three-step specific arguments (\code{assignment}, \code{correction}) can also be passed. A \pkg{StepMix} model for three-step estimation with a BCH correction and soft assignment of a model with categorical indicators \code{Y} and continuous distal outcomes \code{Z\_o} would look like:
\begin{minted}{python}
model = StepMix(
    n_components = 3, 
    measurement = "categorical", 
    structural = "gaussian_unit",
    n_steps = 3, 
    assignment = "soft", 
    correction = "BCH",
)
model.fit(Y, Z_o)
\end{minted}

As with any \pkg{scikit-learn} estimator, \pkg{StepMix} objects come with an array of useful methods for inference and model interpretation:
\begin{itemize}
\item \code{model.predict(Y, Z_o)} to predict the latent class $x$ for each observation;
\item \code{model.predict_proba(Y, Z_o)} to predict $p(x|y, z_o)$ for each observation;
\item \code{model.score(Y, Z_o)} to compute the average log-likelihood of the CM over the dataset;
\item \code{model.aic(Y, Z_o)} to compute the Akaike information criterion~\citep{akaike1974new} over the dataset;
\item \code{model.bic(Y, Z_o)} to compute the Bayesian information criterion~\citep{schwarz1978estimating} over the dataset;
\item \code{model.get_mm_df()} to get the MM model parameters $\theta_Y$ as a data frame;
\item \code{model.get_sm_df()} to get the SM model parameters $\theta_S$ as a data frame;
\item \code{model.get_cw_df()} to get the marginal distribution over latent classes $ p(x ; \theta_X)$ as a data frame;
\item \code{model.sample(100)} to sample 100 observations from the fitted model.
\end{itemize}


The methods can be similarly used with some covariate data \code{Z_p} or some complete model data \code{Z}. We note that sampling from a fitted \pkg{StepMix} model is only possible if the marginal over the latent variable $X$ is explicitly specified, i.e., when no covariates are used in the model (see Equation~\ref{eq:probasX}).

\subsection{Nonparametric bootstrapping}
\label{subsec:bootstrap}
Fitted \pkg{StepMix} objects include a \code{bootstrap_stats} method to obtain mean and standard error estimates of model parameters via nonparametric boostrapping. We run a permutation test to align classes with the main estimator and avoid label switching between repetitions. The method can be called with

\begin{minted}{python}
stats_dict = model.bootstrap_stats(
    Y, 
    Z_o, 
    n_repetitions = 10, 
    progress_bar = True,
)
\end{minted}

which returns a \proglang{Python} dictionary with the following attributes
\begin{itemize}
    \item \code{stats_dict["samples"]}:   Bootstrapped samples in a long-form format;
    \item \code{stats_dict["mm_mean"]}: Means of the MM parameters;
    \item \code{stats_dict["mm_std"]}: Standard errors of the MM parameters;
    \item \code{stats_dict["sm_mean"]}: Means of the SM parameters;
    \item \code{stats_dict["sm_std"]}: Standard errors of the SM parameters;
    \item \code{stats_dict["cw_mean"]}: Means of the class weights (marginal distribution);
    \item \code{stats_dict["cw_std"]}: Standard errors of the class weights.
\end{itemize}

All attributes in the \code{bootstrap_stats} output dictionary are \pkg{Pandas} data frames.

Users can pass the data frame of bootstrapped samples to the \pkg{seaborn} library \cite{Waskom2021} for visualization, as shown in one of our tutorials \footnote{\url{https://colab.research.google.com/drive/14DJCqFTUaYp3JtLAeAMYmGHFLCHE-r7z}}.

\subsection{Additional features}
\pkg{StepMix} provides additional functionalities, which are detailed in the documentation and tutorials, including:
\begin{itemize}
  \item Support for \textbf{missing values}. Most models introduced in Table~\ref{tab:StepMix_models} can be suffixed with \code{\_nan} (e.g., \code{binary\_nan}) to enable full information maximum likelihood training (Section~\ref{Subsec:Likelihood}). The \code{gaussian\_full} and \code{covariate} models do not currently support FIML;
  \item A dictionary-based syntax for combining multiple models in \code{measurement} or \code{structural}, allowing, for example, to declare a MM with \textbf{both categorical and continuous} variables or a SM with a covariate and an outcome (see example in Section~\ref{Subsec:complete_simulation}).
\end{itemize}

Finally, \pkg{StepMix} was designed with modular methods for stepwise estimation, meaning each step in the estimation procedures can be represented as a standalone method call. For example, the naive three-step method can also be used by independently calling methods for i) fitting the MM with the EM algorithm, ii) performing soft assignments, and iii) fitting the SM by calling the M-step:

\begin{minted}{python}
model.em(Y)                        
soft_assignments = model.predict_proba(Y)      
model.m_step_structural(soft_assignments, Z_o) 
\end{minted}

Access to lower-level steps in the estimation will hopefully further stepwise estimation research by allowing researchers to investigate existing methods and more easily implement new ones.




\section{Computational examples}
\label{Sec:Computational_examples}
In this section, we carry out two simulation studies to replicate \pkg{Latent Gold} results presented in \citet{bakk2018two} (Sections~\ref{Subsec:response_simulation} and \ref{Subsec:response_simulation}) followed by another simulation to test a complete model in the presence of missing values (Section~\ref{Subsec:complete_simulation}). We then present a case study on a real-life dataset (Section~\ref{Subsec:real_data}) and conclude with a comparison of \pkg{StepMix} log-likelihoods with other packages on various datasets (Section~\ref{Subsec:comparison_packages}). 

The \pkg{StepMix} source code includes functions to simulate the datasets used in this section as well as instructions to reproduce our results. See the \code{scripts} directory at \url{https://github.com/Labo-Lacourse/stepmix}.

\subsection{Distal outcome simulation study}
\label{Subsec:response_simulation}
For the first simulation study, we consider a dataset with $K=3$ latent classes, $D_{M} = 6$ binary indicator variables, and $D_{S}=1$ continuous outcome variable, which is modeled as a mixture of unit variance Gaussians (see Table~\ref{tab:StepMix_models}). Therefore, the MM's parameters are given by $\theta_M = \{\rho, \ \pi\}$, where $\pi \in [0, 1]^{D_{M} \times K}$ represents class-conditioned binary probabilities, and the marginal distribution of the latent variable respects $p(X{=}k) = \rho_k$. The SM's parameters are $\theta_S = \{\mu\}, \ \mu \in \mathbb{R}^{D_{S} \times K}$. The true values of the parameters are:
\begin{align*}
\begin{cases}
\rho &= [1/3, \ 1/3, \ 1/3] , \hspace{0.5cm}\mu = [-1, \ 1, \ 0],\\ \ \\
\pi &= \begin{bmatrix}
    \gamma       & \gamma & 1 - \gamma \\
    \gamma       & \gamma & 1 - \gamma  \\
    \gamma       & \gamma & 1 - \gamma \\ 
    \gamma       & 1 - \gamma & 1 - \gamma \\
    \gamma       &  1 - \gamma & 1 - \gamma  \\
    \gamma       &  1 - \gamma & 1 - \gamma
\end{bmatrix},
\end{cases}
\end{align*}
where the parameter $\gamma$ controls the strength of the association between the latent class and the binary variables. \cite{bakk2018two} use for $\gamma$ the values 0.9, 0.8, and 0.7 and call them the high-, medium-, and low-separation
conditions, respectively.

To respect the original experiment design, we consider simulations with sample sizes $n$ of 500,
1000, and 2000, resulting in nine configurations (three sample sizes $\times$ three class separations). For each configuration, 500 data sets are generated. For simplicity, the authors only reported results for the estimates of parameter $\mu_2 = 1$. We therefore present the same statistics for all estimation methods. It should be noted that we use modal assignments for all three-step simulations following the specifications in \citet{bakk2018two}. To minimize the variance in the estimation of the bias and the root-mean-squared error (RMSE) of the different estimators of $\mu_2$ in different class separation settings, we used common random numbers to generate our simulated datasets. 

The \pkg{StepMix} code for a single simulation of the high-separation one-step case is the following. Please note that the distal outcome is referred to as the \code{response} in the source code due to the terminology used in \cite{bakk2018two}. We begin by simulating the dataset---\code{Y} is the matrix of MM data and \code{Z_o} is the matrix of SM outcome data--- before instantiating the model and fitting it. We finally retrieve the means of the SM.

\begin{minted}{python}
from stepmix.datasets import data_bakk_response
from stepmix.stepmix import StepMix

Y, Z_o, _ = data_bakk_response(
    n_samples = 2000,
    sep_level = 0.9,
    random_state = 42,
)

model = StepMix(
    n_components = 3,
    measurement = "binary",
    structural = "gaussian_unit",
    n_steps = 1,
    random_state = 42,
    verbose = 1,
)
model.fit(Y, Z_o)

mus = model.get_sm_df()
\end{minted}

\input{tables/sim_response}

 The model report for the above example is presented in Appendix~\ref{subsec:sim_output}. 
 
The full results of our simulations are shown in Table~\ref{tab:sim}. The \pkg{StepMix} results are consistent with the ones presented in Table~\ref{tab:Bakk}, taken from \citet{bakk2018two} in which models were estimated with \pkg{Latent Gold} version 5.1. The most noticeable difference is that our implementations of the two-step and ML methods seem to produce slightly better estimates compared to the commercial implementation in the low-separation condition, especially with small sample sizes. 

\input{tables/sim_response_bakk}

The general conclusions presented by \citet{bakk2018two} can be drawn from both Tables~\ref{tab:sim}~and~\ref{tab:Bakk}. The one-step estimator is globally unbiased and systematically has the lowest RMSE. The naive three-step estimator is severely biased and has the highest RMSE. Its bias decreases with increasing class separation and is essentially unaffected by sample size. As the class separation level and the sample size increase, the bias-adjusted three-step methods progressively reduce this bias.

The two-step estimator is consistently better than the bias-adjusted three-step estimators. Its smaller RMSE suggests that avoiding the extra step of three-step estimation may improve the quality of the resulting estimator. This conclusion appears to be quite intuitive since the second step of the three-step methods does not add any information to the model that was not already encapsulated in the MLE $\hat{\theta}_{M}$ obtained in step one. In the medium- and high-separation conditions, the two-step estimator also performs essentially as well as the one-step estimator, suggesting that there is little loss of efficiency from moving from one-step estimation to an appropriate multi-step approach in this setting.

\subsection{Covariate simulation study}
\label{Subsec:covariate_simulation}
The second simulation is concerned with a structural covariate model where the latent class $X$ is a response to an observed one-dimensional predictor $Z^p$. Specifically, the SM is a multinomial logistic model (Table~\ref{tab:StepMix_models}, \code{covariate}). 
 For the simulation at hand, $Z^p$ is a uniformly sampled integer between 1 and 5. The ground truth parameters of the structural model are:
\begin{align*}
\begin{cases}
\beta &= [0.00, \ -1.00,\ 1.00], \\
b &= [0.00, \ 2.35, \ -3.66], \\
\end{cases}
\end{align*}
with $\theta_S = (\beta, b)$. The intercepts $b$ were tuned to ensure approximately equal class sizes when averaged over $Z^p$. We consider the same binary indicators $\pi$ as in the outcome simulation (Section~\ref{Subsec:response_simulation}), and the parameters of the MM are now defined as $\theta_M = (\pi)$, since we ignore the class weights $\rho$ to follow the conditional likelihood perspective presented in Equation~\ref{eq:probasX} for the covariate case. The simulation specifications are otherwise identical to the ones presented in Section~\ref{Subsec:response_simulation}. 

Similarly to Section~\ref{Subsec:response_simulation}, we present the code for a single covariate simulation. We begin by simulating the dataset.
\begin{minted}{python}
from stepmix.datasets import data_bakk_covariate
from stepmix.stepmix import StepMix

Y, Z_p, _ = data_bakk_covariate(
    n_samples = 2000,
    sep_level = 0.9,
    random_state = 42,
)
\end{minted}

Contrary to other models, the covariate model does not have a closed-form MLE for the M-Step and requires a numerical optimization method for fitting parameters, the arguments of which can be optionally specified by passing a \proglang{Python} dictionary to the \code{structural\_params} parameter in \pkg{StepMix}.
\begin{minted}{python}
covariate_params = {
    "method": "newton-raphson",
    "max_iter": 1,
    "intercept": True,
}

model = StepMix(
    n_components = 3,
    measurement = "binary",
    structural = "covariate",
    n_steps = 1,
    random_state = 42,
    structural_params = covariate_params,
)
model.fit(Y, Z_p)
\end{minted}

We conclude by retrieving the coefficients of the SM and using the \pkg{Pandas} library API to subtract the coefficients of the second class from all coefficients, therefore setting this class as the reference class.

\begin{minted}{python}
betas = model.get_sm_df()

betas = betas.sub(betas[1], axis = 0)
\end{minted}

Covariate simulation results are presented in Table~\ref{tab:sim_covariate}. The results largely echo the ones presented in Section~\ref{Subsec:response_simulation}, with \pkg{StepMix} yielding by and large similar estimates to those found in \citet{bakk2018two} and reproduced in Table~\ref{tab:Bakk_covariate}. Again, stepwise estimators (except naive three-step) produce globally unbiased estimates, and the \pkg{StepMix} implementation appears to slightly outperform \pkg{Latent Gold} in the two-step and ML cases.

\input{tables/sim_covariate}
\input{tables/sim_covariate_bakk}

\subsection{Complete simulation study}
\label{Subsec:complete_simulation}
As a final simulation, we merge the outcome and covariate models of the previous sections into a CM. We fix the degree of class separation in the MM ($\gamma=0.8$) and instead vary the ratio of data missing completely at random in both the MM and the outcome to showcase our FIML implementation. 

We again begin our example by simulating the dataset. This time, \code{Z} includes the covariate variable as the first column and the outcome as the second column.
\begin{minted}{python}
from stepmix.datasets import data_bakk_complete
from stepmix import StepMix

Y, Z, _ = data_bakk_complete(
    n_samples = 2000,
    sep_level = 0.8,
    nan_ratio = 0.25,
    random_state = 42,
)
\end{minted}

While we previously specified models as simple strings (e.g., \code{"binary"}), the structural model for this example requires both a covariate and a continuous outcome. This setting requires a more complex model description, which with \pkg{StepMix} can be achieved using a nested \proglang{Python} dictionary. The nested dictionaries represent different submodels using different columns of \code{Z} as inputs. The keys in the main dictionary can be arbitrary strings to name the submodels. 

The following descriptor indicates that the first column in \code{Z} should be passed to a \code{covariate} submodel with the given \code{method} and \code{max_iter} arguments. The next column should be passed to a \code{gaussian_unit_nan} submodel. The \code{_nan} suffix ensures support for missing values.

\begin{minted}{python}
structural_descriptor = {
    "covariate": {
        "model": "covariate",
        "n_columns": 1,
        "method": "newton-raphson",
        "max_iter": 1,
    },
    "response": {
        "model": "gaussian_unit_nan",
        "n_columns": 1,
    },
}
\end{minted}

The above descriptor can be extended to accommodate an arbitrary number of columns and submodels. We could also define a similar \code{measurement_descriptor}, although this is not needed here since the MM is homogenous and only includes \code{binary} variables. We conclude this example by instantiating a \pkg{StepMix} estimator and fitting it. Now that the structural model is more complex, we rely on the \pkg{Pandas} API to specifically select the parameters of the SM model called \code{response}, as defined by the key in the descriptor.

\begin{minted}{python}
model = StepMix(
    n_components = 3,
    measurement = "binary_nan",
    structural = structural_descriptor,
    n_steps = 1,
    random_state = 42,
)
model.fit(Y, Z)

mus = model.get_sm_df().loc["response"]
\end{minted}

We present the complete simulation results in Table~\ref{tab:Bakk_complete}. We find that the one-step, two-step, and ML methods are globally unbiased, even in the presence of data missing completely at random. As expected, the RMSE increases with the ratio of missing values due to the decreasing effective sample size associated with each parameter. Naive three-step (and, to a lesser extent, three-step with BCH correction) yields noticeably inferior estimates as the ratio of missing data increases.

\input{tables/sim_complete}
\newpage
\subsection{Application example: Using parents' social status to predict respondents' family income}
\label{Subsec:real_data}

In this section, we illustrate the use of \pkg{StepMix} on a real-life dataset ($n$=3029) that was previously studied by \cite{bakk2016relating} and \cite{bakk2021relating}. This analysis is based on the 1976 ($n_1$=1499) and 1977 ($n_2$=1530) American General Social Surveys (GSS) conducted by the National Opinion Research Center. The data is publicly accessible via the \pkg{gssr} package \citep{gssr}.

We consider a MM relating three indicators (the father’s job prestige, the mother's highest degree and the father's highest degree) to a latent variable with $K=3$ latent classes, interpreted as the parents’ social status (low, middle and high). The father’s job prestige score was measured on a scale ranging from 12 to 82 and recoded into three categories: low (36 or less), medium (37-61) and high (62 or above). The father’s and mother's education were measured on a 5-point scale ranging from 0 (less than high school) to 4 (graduate). In the SM, the parents’ social status is used to predict the respondent's annual family income, in thousands of dollars. To account for the skewed distribution of this unique distal outcome, it is modeled as a normal variable with class-dependent mean and variance. Missing values for the annual family income are present in the GSS dataset.

This model was first estimated by \cite{bakk2016relating} using the stepwise estimation methods implemented in \pkg{Latent Gold} and handling the missing data through listwise deletion. \cite{bakk2021relating} also compared the one-, two-, and three-step (BCH and ML) approaches on this dataset, but this time using FIML. We apply the same experimental setup as the latter study, but using \pkg{StepMix} instead of \pkg{Latent Gold} for the estimation of each model. Appendix~\ref{subsec:real_output} includes an example of a \pkg{StepMix} ML model report after fitting the GSS data.

Table~\ref{tab:Real_data_MM} presents the estimated class proportions and the resulting class-conditional empirical distribution of the MM's indicators. Table~\ref{tab:Real_data_SM} reports the estimated parameters of the SM provided by each stepwise approach and their standard errors (SEs) and $p$~values, estimated by nonparametric bootstrapping based on 100 repetitions\footnote{The reported SEs are obtained by calculating the standard deviation (SD) of a given bootstrapped parameter using StepMix’s nonparametric bootstrap module (Section~\ref{subsec:bootstrap}). These bootstrapped parameters are the mean incomes for each latent class in Table~\ref{tab:Real_data_SM} and the difference between the mean income of the reference latent class (low) and those of the middle and high classes in Table~\ref{tab:Real_data_SM_anova}.}. Table~\ref{tab:Real_data_SM_anova} presents two-tailed $Z$-tests to assert whether the class-specific mean incomes for the middle and high classes are identical to that of the low social class.

\input{tables/Real_data_table_MM}

\input{tables/Real_data_table_SM}

\input{tables/Real_data_table_SM_anova}

The results obtained with \pkg{StepMix} are consistent with those reported by \cite{bakk2021relating}. Table~\ref{tab:Real_data_MM} indicates that the low, middle and high social classes respectively compose approximately 70\%, 23\% and 7\% of the sample. The class-conditional distribution of the indicators is highly heterogeneous, which highlights a strong association between the level of education and job prestige of the parents. 

The results of Table~\ref{tab:Real_data_SM} show that the estimated SM's parameters vary considerably between methods. For example, the estimated average family income in the highest social class ranges from 43.67 thousand dollars a year with the naive three-step method to 67.90 thousand dollars a year with the one-step method. In this case study, the assumption of normality of the class-conditional annual income is violated, as previously discussed by \cite{bakk2016relating} and \cite{bakk2021relating}. In the one-step approach, the joint estimation of the MM and the SM propagates this specification error from the SM to the definition of the latent classes. As a result, the estimated proportion of each class in the sample changes to 62\%, 20\% and 18\%, and the empirical distribution of the indicators in both smaller classes is practically indistinguishable. This observation illustrates that the definition of the SM can significantly impact the interpretation of the latent class in the one-step approach, making it less suitable for real-life applications.

Regarding the variance of the estimators, the naive three-step method leads to the lowest bootstrap estimates of the SM's parameters' SEs. However, the resulting estimated class-conditional average incomes are biased since this method does not consider the uncertainty of class assignments. Among the two-step and bias-corrected three-step estimators, the BCH method has the smallest SEs across bootstrapped means. This is consistent with the results obtained by \cite{bakk2021relating}, whose results indicate that BCH method is the least sensitive to violations of distributional assumptions. Compared to the three-step ML approach, the two-step method has noticeably closer estimated means to the BCH method.

These differences in the estimated SM's parameters and their SEs are important from an applied perspective as they can lead to significant differences in the conclusion of the study. Based on the $p$~values of the $Z$-tests of Table~\ref{tab:Real_data_SM_anova}, researchers would reach different conclusions regarding the difference in respondent’s family income between those whose parents were classified in the low and middle social statuses. According to the more robust BCH approach, we should conclude that the incomes are, on average, significantly higher when their parents had a middle social status ($Z$= 5.24; $p < 0.001$) than low social status. Researchers using the one-step ($Z$ = 1.53; $p = 0.126$) or the two-step approach ($Z$ = 1.57; $p = 0.116$) would incorrectly fail to reject the null hypothesis. This highlights the importance of comparing different stepwise estimators when studying real data, as imperfect model specification is almost unavoidable in practice.

\subsection{Comparison with other packages}
\label{Subsec:comparison_packages}
\input{tables/comparison_packages}

We present a comparison of the log-likelihoods obtained with \pkg{StepMix} and four other mixture modeling packages on six datasets in Table~\ref{tab:comparison_packages}. \pkg{StepMix} yields log-likelihoods that are nearly identical to existing commercial and open-source software, validating the quality of our implementation.

\section{Conclusion}
\label{Sec:Conclusion}
We presented \pkg{StepMix}, an open-source package for mixture model estimation with continuous and categorical variables. To the best of our knowledge, \pkg{StepMix} is the first \proglang{Python} package to implement bias-adjusted stepwise estimators for mixture models with external variables, potentially exposing a completely new user base to these methods. The familiar \pkg{scikit-learn} interface, the support for missing values, and the \proglang{R} wrapper are additional features that we believe will make \pkg{StepMix} a relevant tool for the community. Furthermore, we presented all estimation procedures as variants of the EM algorithm to provide a unified framework of analysis. This framework also guided the implementation of \pkg{StepMix}, resulting in modular source code that is conducive to the implementation of new features, such as novel stepwise estimation methods or additional conditional distributions. As a future development, all of the 14 Gaussian covariance decompositions implemented in \pkg{mclust} and \pkg{MoEClust} could be added to \pkg{StepMix}. We could also implement efficient initialization methods for the EM algorithm instead of relying on random initialization. It would also be interesting to extend the capabilities of \pkg{StepMix} to deal with high dimensional data using recent regularization techniques \citep{robitzsch2020regularized}, and potentially handle large data sets through the implementation of additional EM methods. Other extensions could include the support for generalized mixture models for longitudinal data, such as growth mixture models \citep{ram2009methods}, latent class growth models \citep{nagin1999analyzing}, and mixtures of hidden Markov models \citep{van1990mixed}. 

\section{Acknowledgments}

This research was funded in part by Natural Sciences and Engineering Research Council of Canada (NSERC) PGS D Scholarships [S.M., R.L.], by Fonds de recherche du Québec - Nature et technologies (FRQNT) master's [S.M., R.L.] and doctoral scholarships [S.M., R.L.], by IVADO MSc Excellence Scholarships [S.M., R.L.], as well as by the Canadian Institutes of Health Research (CIHR) (grant number 170633), the Social Sciences and Humanities Research Council (SSHRC) and the Centre for the Study of Democratic Citizenship (CSDC) [R.S.]. This research was enabled in part by computing resources provided by Mila (mila.quebec).

\newpage

\bibliography{references}
\newpage
\appendix
\section{Appendix}
\label{Sec:Appendix}
\setminted{fontsize=\small}
\subsection{StepMix output of the distal outcome simulation study}
\label{subsec:sim_output}
By setting \code{verbose=1}, \pkg{StepMix} outputs a full model report, including parameter estimates and fit statistics. The output for the code example in Section~\ref{Subsec:response_simulation} is as follows.
\setminted{fontsize=\small}
\begin{minted}{text}
Fitting StepMix...
================================================================================
MODEL REPORT
================================================================================
    ============================================================================
    Measurement model parameters
    ============================================================================
          model_name       binary                
          class_no              0       1       2
          param variable                         
          pis   feature_0  0.9028  0.0928  0.9024
                feature_1  0.8996  0.1044  0.9108
                feature_2  0.8978  0.1024  0.9039
                feature_3  0.1156  0.0832  0.8992
                feature_4  0.0756  0.0885  0.9034
                feature_5  0.1004  0.0962  0.9080
    ============================================================================
    Structural model parameters
    ============================================================================
          model_name      gaussian_unit                
          class_no                    0       1       2
          param variable                               
          means feature_0        0.9599 -0.0379 -1.0153
    ============================================================================
    Class weights
    ============================================================================
        Class 1 : 0.34
        Class 2 : 0.34
        Class 3 : 0.32
    ============================================================================
    Fit for 3 latent classes
    ============================================================================
    Estimation method             : 1-step
    Number of observations        : 2000
    Number of latent classes      : 3
    Number of estimated parameters: 23
    Log-likelihood (LL)           : -8600.7930
    -2LL                          : 17201.5860
    Average LL                    : -4.3004
    AIC                           : 17247.59
    BIC                           : 17376.41
    CAIC                          : 17399.41
    Sample-Size Adjusted BIC      : 17478.16
    Entropy                       : 183.3425
    Scaled Relative Entropy       : 0.9166

\end{minted}

\subsection{StepMix output of the application example}
\label{subsec:real_output}
\pkg{StepMix} output for the ML estimator on the GSS data presented in Section~\ref{Subsec:real_data}.

\setminted{fontsize=\small}
\begin{minted}{text}
Fitting StepMix...
================================================================================
MODEL REPORT
================================================================================
    ============================================================================
    Measurement model parameters
    ============================================================================
          model_name                    categorical_nan                
          class_no                                    0       1       2
          param variable                                               
          pis   Father's education_0             0.9452  0.0553  0.0000
                Father's education_1             0.0533  0.8863  0.1137
                Father's education_2             0.0015  0.0050  0.0530
                Father's education_3             0.0000  0.0534  0.3899
                Father's education_4             0.0000  0.0000  0.4434
                Father's job prestige_0          0.4667  0.3056  0.0497
                Father's job prestige_1          0.5304  0.6735  0.4616
                Father's job prestige_2          0.0029  0.0209  0.4887
                Mother's education_0             0.8215  0.1376  0.1523
                Mother's education_1             0.1658  0.7857  0.4352
                Mother's education_2             0.0039  0.0295  0.0107
                Mother's education_3             0.0076  0.0399  0.3012
                Mother's education_4             0.0012  0.0073  0.1006
    ============================================================================
    Structural model parameters
    ============================================================================
          model_name                gaussian_diag_nan                     
          class_no                                  0         1          2
          param       variable                                            
          covariances Income (1000)          144.7759  607.0978  2633.7187
          means       Income (1000)           20.7196   45.2885    66.0730
    ============================================================================
    Class weights
    ============================================================================
        Class 1 : 0.70
        Class 2 : 0.23
        Class 3 : 0.07
    ============================================================================
    Fit for 3 latent classes
    ============================================================================
    Estimation method             : 3-step
    Correction method             : ML
    Assignment method             : modal
    Number of observations        : 3029
    Number of latent classes      : 3
    Number of estimated parameters: 38
    Log-likelihood (LL)           : -18754.7237
    -2LL                          : 37509.4475
    Average LL                    : -6.1917
    AIC                           : 37585.45
    BIC                           : 37814.05
    CAIC                          : 37852.05
    Sample-Size Adjusted BIC      : 37997.92
    Entropy                       : 624.9492
    Scaled Relative Entropy       : 0.8122
\end{minted}

\subsection[]{Distal outcome simulation in \proglang{R}}
\label{subsec:R_sim1}
This section replicates the \proglang{Python} example from section \ref{Subsec:response_simulation} using the \proglang{R} interface \pkg{StepMixR}. 
\begin{minted}{R}
library(stepmixr)

datasim = data_bakk_response(
  n_samples = 2000, 
  sep_level = 0.9, 
  random_state = 42
  )

Y = datasim[[1]]
Z_o = datasim[[2]]

model = stepmix(
  n_components = 3, 
  measurement = 'binary', 
  structural = 'gaussian_unit', 
  n_steps = 1, 
  random_state = 42, 
  verbose = 1
  ) 

fit1 = fit(model, Y, Z_o)

mus = fit1$get_sm_df()
\end{minted}

\subsection[]{Covariate simulation in \proglang{R}}
\label{subsec:R_sim2}
This section replicates the \proglang{Python} example from section \ref{Subsec:covariate_simulation} using the \proglang{R} interface \pkg{StepMixR}. 

\begin{minted}{R}
library(stepmixr)

datasim = data_bakk_covariate(
  n_samples = 2000, 
  sep_level = 0.9, 
  random_state = 42
  )

Y = datasim[[1]]
Z_p = datasim[[2]]

covariate_params = list(
  method = 'newton-raphson', 
  max_iter = as.integer(1), 
  intercept = TRUE
  )

model = stepmix(
  n_components = 3, 
  measurement = 'binary', 
  structural = 'covariate', 
  n_steps = 1, 
  random_state = 42, 
  structural_params = covariate_params
  )

fit1 = fit(model, Y, Z_p)

betas = fit1$get_parameters()

betas = betas$structural$beta

betas = betas - c(1, 1, 1) %o% betas[2, ]
\end{minted}

\subsection[]{Complete simulation study in \proglang{R}}
\label{subsec:R_sim3}
This section replicates the \proglang{Python} example from section \ref{Subsec:complete_simulation} using the \proglang{R} interface \pkg{StepMixR}. 

\begin{minted}{R}
library(stepmixr)

datasim = data_bakk_complete(
  n_samples = 2000, 
  sep_level = 0.8, 
  nan_ratio = 0.25, 
  random_state = 42
  )

Y = datasim[[1]]
Z = datasim[[2]]

structural_descriptor = list(
  covariate = list( 
    model = "covariate", 
    n_columns = as.integer(1), 
    method = "newton-raphson", 
    max_iter = as.integer(1)
  ),
  response = list(
    model = "gaussian_unit_nan", 
    n_columns = as.integer(1)
  )
)

model = stepmix(
  n_components = 3, 
  measurement = "binary_nan", 
  structural = structural_descriptor, 
  n_steps = 1, 
  random_state = 42
  )

fit1 = fit(model, Y, Z)

mus = fit1$get_parameters()
mus = mus$structural$response
\end{minted}

\end{document}

%% file: figures/fig_complete_model.tex
\begin{figure}[H]
 \begin{center}
     \begin{tikzpicture}[scale=1.7]
     \node[draw,circle, minimum size=1.0cm] (X) at (1,1) { $\textcolor{black}{X}$ };
      \node[draw,circle,fill=lightgray, minimum size=1.0cm] (Y) at (1,0) {$Y$};
      \node[draw,circle,fill=lightgray, minimum size=1.0cm] (Zc) at (0,1) { $Z^p$ };
      \node[draw,circle,fill=lightgray, minimum size=1.0cm] (Zo) at (2,1) { $Z^o$ };
      \draw[>=latex,->,color=black](X.south) -- (Y.north);
      \draw[>=latex,->,color=black](Zc.east) -- (X.west);
      \draw[>=latex,->,color=black](X.east) -- (Zo.west);
      \end{tikzpicture}
\end{center}
\caption{Family of mixture models that can be estimated by \textbf{StepMix}. Conditioning on the latent class $X$ blocks the paths between the observed indicators, illustrating graphically the conditional independence assumptions and how the joint likelihood factorizes (Equation~\ref{eq:probasCM}). }
\label{fig:CM}
\end{figure}
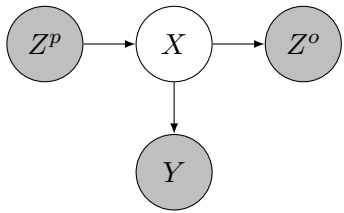

%% file: tables/package_comparison.tex
\newcommand{\yes}{\checkmark}
\newcommand{\No}{}  
\begin{table}[t!]
\begin{center}
\resizebox{\columnwidth}{!}{%
    \begin{tabular}{lcccccccc}
    \hline
        Package & Version & \proglang{R} & \proglang{Python} & \specialcell{\pkg{scikit-learn} \\ API} & \specialcell{Two-step\\ estimation \\} & \specialcell{Bias-adjusted \\ three-step \\ estimation \\} & \specialcell{Gaussian and \\ non-Gaussian \\ components}  & \specialcell{Covariates} \\ \hline
        \pkg{StepMix} & 2.1.3 & \yes & \yes & \textbf{\yes} & \yes & \yes & \yes & \yes \\ \hline
        \pkg{scikit-learn} & 1.2 & \No & \yes & \textbf{\yes} & \No & \No & \No & \No \\ \hline
        \pkg{multilevLCA} & 1.1 & \yes & \No & \No & \yes & \No & \yes & \yes \\ \hline
        \pkg{mclust} & 6.0 & \yes & \No & \No & \No & \No & \No & \No \\ \hline
        \pkg{MoEClust} & 1.5.2 & \yes & \No & \No & \No & \No & \No & \yes \\ \hline
        \pkg{AutoGMM} & 2.0.1 & \No & \yes & \yes & \No & \No & \No & \No \\ \hline
        \pkg{MixMod} & 0.2.0 & \No & \yes & \No & \No & \No & \yes & \No \\ \hline
        \pkg{Rmixmod} & 2.1.8 & \yes & \No & \No & \No & \No & \yes & \No \\ \hline
        \pkg{poLCA} & 1.6.0.1 & \yes & \No & \No & \No & \No & \No & \yes \\ \hline
        \pkg{depmixS4} & 1.5 & \yes & \No & \No & \No & \No & \yes & \yes \\ \hline
        \pkg{randomLCA} & 1.1-2 & \yes & \No & \No & \No & \No & \No & \No \\ \hline
        \pkg{BayesLCA} & 1.9 & \yes & \No & \No & \No & \No & \No & \No \\ \hline
        \pkg{e1071::lca} & 1.7-13 & \yes & \No & \No & \No & \No & \No & \No \\ \hline
        \pkg{glca} & 1.3.3 & \yes & \No & \No & \No & \No & \No & \yes \\ \hline
        \pkg{VarSelLCM} & 2.1.3.1 & \yes & \No & \No & \No & \No & \yes & \No \\ \hline
        \pkg{FlexMix} & 2.3-18 & \yes & \No & \No & \No & \No & \yes & \yes \\ \hline
\end{tabular}}
\caption{Features of open-source software packages for mixture models estimation.}
\end{center}
\label{tab:packagecomparison}
\end{table}

%% file: pseudocode/1-step.tex
\begin{algorithm}[H]
\caption{One-step method}\label{alg:1-step}
\begin{algorithmic}

\State Initialize $(\hat{\theta}_M^{(0)}, \hat{\theta}_S^{(0)}) \in (\Theta_M \times \Theta_S)$\\
\vspace{-0.2cm}

\hrulefill
\State \underline{\textbf{STEP 1 : EM algorithm on the CM}}

\State Set $t=0$

\Repeat
\vspace{-0.1cm}

    \State \hrulefill \text{ E Step }\hrulefill
    
    \State $\tau^{(t)}_{ik} = p(X_i{=}k|y_i,z^o_i;z^p_i,\hat{\theta}_M^{(t)}, \hat{\theta}_S^{(t)} ), \ \forall i \in N, \forall k \in \mathcal{X}$ \Comment{Update responsibilities}
    \vspace{0.2cm}

    \State \hrulefill \text{ M Step }\hrulefill

    \State $\hat{\theta}_{M}^{(t+1)} = \argmax\limits_{\theta_{M} \in \Theta_{M}}\sum\limits_{i\in N}\omega_i\sum\limits_{k\in\mathcal{X}} \tau^{(t)}_{ik}\log p(X_i{=}k, y_i;\theta_{M})$
    \Comment{Update $\hat{\theta}_{M}$}

    \State $\hat{\theta}_{S}^{(t+1)} = \argmax\limits_{\theta_{S} \in \Theta_{S}}\sum\limits_{i\in N}\omega_i\sum\limits_{k\in\mathcal{X}} \tau^{(t)}_{ik}\log p(X_i{=}k, z^o_i;z^p_i,\theta_{S})$
    \Comment{Update $\hat{\theta}_{S}$}

    \State $t=t+1$
\Until{convergence}

\State Set $(\hat{\theta}_M, \hat{\theta}_S) =(\hat{\theta}_M^{(t)}, \hat{\theta}_S^{(t)})$ \Comment{Estimated CM parameters}

\\\hrulefill
\State Return $(\hat{\theta}_M, \hat{\theta}_S)$ \Comment{Return estimated CM parameters}
\end{algorithmic}
\end{algorithm}

%% file: pseudocode/2-step.tex
\begin{algorithm}[H]
\caption{Two-step method}\label{alg:2-step}
\begin{algorithmic}

\State Initialize $(\hat{\theta}_M^{(0)}, \hat{\theta}_S^{(0)}) \in (\Theta_M \times \Theta_S)$\\
\vspace{-0.2cm}

\hrulefill
\State \underline{\textbf{STEP 1 : EM algorithm on the MM}}

\State Set $t=0$

\Repeat
\vspace{-0.2cm}

    \State \hrulefill \text{ E Step }\hrulefill
    
    \State $\tau^{(t)}_{ik} = p(X_i{=}k|y_i;\hat{\theta}_M^{(t)}), \ \forall i \in N_1, \forall k \in \mathcal{X}$ \Comment{Update responsibilities}

    \State \hrulefill \text{ M Step }\hrulefill

    \State $\hat{\theta}_{M}^{(t+1)} = \argmax\limits_{\theta_{M} \in \Theta_{M}}\sum\limits_{i\in N_1}\omega_i\sum\limits_{k\in\mathcal{X}} \tau^{(t)}_{ik}\log p(X_i{=}k, y_i;\theta_{M})$
    \Comment{Update $\hat{\theta}_{M}$}

    \State $t=t+1$
\Until{convergence}

\State Set $\hat{\theta}_M = \hat{\theta}_M^{(t)}$ \Comment{Estimated MM parameters}

\\\hrulefill
\State \underline{\textbf{STEP 2 : EM algorithm on the CM with $\hat{\theta}_M$ fixed}}

\State Set $t=0$

\Repeat
\vspace{-0.2cm}

    \State \hrulefill \text{ E Step }\hrulefill
    
    \State $\tau^{(t)}_{jk} = p(X_j{=}k|y_j,z^o_j;z^p_j,\hat{\theta}_M, \hat{\theta}_S^{(t)} ), \ \forall j \in N_2, \forall k \in \mathcal{X}$ \Comment{Update responsibilities}
    \vspace{0.2cm}

    \State \hrulefill \text{ M Step }\hrulefill

    \State $\hat{\theta}_{S}^{(t+1)} = \argmax\limits_{\theta_{S} \in \Theta_{S}}\sum\limits_{j\in N_2}\omega_j\sum\limits_{k\in\mathcal{X}} \tau^{(t)}_{jk}\log p(X_j{=}k, z^o_j;z^p_j,\theta_{S})$
    \Comment{Update $\hat{\theta}_{S}$}

    \State $t=t+1$
\Until{convergence}

\State Set $\hat{\theta}_S = \hat{\theta}_{S}^{(t)}$ \Comment{Estimated SM parameters}

\\\hrulefill
\State Return $(\hat{\theta}_M, \hat{\theta}_S)$ \Comment{Return estimated CM parameters}
\end{algorithmic}
\end{algorithm}

%% file: figures/fig_structural_model_ML.tex
\begin{figure}[H]
 \begin{center}
     \begin{tikzpicture}[scale=1.7]
     \node[draw,circle, minimum size=1.0cm] (X) at (1,1) { $\textcolor{black}{X}$ };
      \node[draw,circle,fill=lightgray, minimum size=1.0cm] (W) at (1,0) {$W$};
      \node[draw,circle,fill=lightgray, minimum size=1.0cm] (Zc) at (0,1) { $Z^p$ };
      \node[draw,circle,fill=lightgray, minimum size=1.0cm] (Zo) at (2,1) { $Z^o$ };
      \draw[>=latex,->,color=black](X.south) -- (W.north);
      \draw[>=latex,->,color=black](Zc.east) -- (X.west);
      \draw[>=latex,->,color=black](X.east) -- (Zo.west);
      \end{tikzpicture}
\end{center}
\caption{Model used in the third step of the three-step method with ML correction.}
\label{fig:SM_ML}
\end{figure}
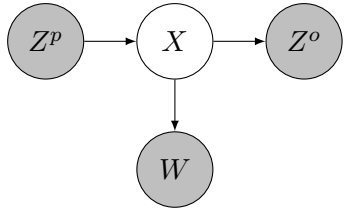

%% file: pseudocode/3-step.tex
\begin{algorithm}[H]
\caption{Three-step method}\label{alg:3-step}
\begin{algorithmic}

\State Initialize $(\hat{\theta}_M^{(0)}, \hat{\theta}_S^{(0)}) \in (\Theta_M \times \Theta_S)$, select $\mathcal{A} \in \{\texttt{soft, modal}\}$, select $\mathcal{C} \in \{\texttt{ML, BCH, none}\}$\\

\hrulefill
\State \underline{\textbf{STEP 1 : EM algorithm on the MM}}

\State Set $t=0$

\Repeat
\vspace{-0.2cm}

    \State \hrulefill \text{ E Step }\hrulefill
    
    \State $\tau^{(t)}_{ik} = p(X_i{=}k|y_i;\hat{\theta}_M^{(t)}), \ \forall i \in N_1, \forall k \in \mathcal{X}$ \Comment{Update responsibilities}

    \State \hrulefill \text{ M Step }\hrulefill

    \State $\hat{\theta}_{M}^{(t+1)} = \argmax\limits_{\theta_{M} \in \Theta_{M}}\sum\limits_{i\in N_1}\omega_i\sum\limits_{k\in\mathcal{X}} \tau^{(t)}_{ik}\log p(X_i{=}k, y_i;\theta_{M})$
    \Comment{Update $\hat{\theta}_{M}$}

    \State $t=t+1$
\Until{convergence}

\State Set $\hat{\theta}_M = \hat{\theta}_M^{(t)}$ \Comment{Estimated MM parameters}

\vspace{-0.1cm}
\\\hrulefill
\State \underline{\textbf{STEP 2 : Class weights based on the MM}}

\State Set $w_{jk}=\begin{cases}
    p(X_j{=}k|y_j;\hat{\theta}_M) & \text{, if $\mathcal{A} = \texttt{soft}$}\\
    \mathds{1}[\argmax_{c \in \mathcal{X}}p(X_j{=}c|y_j;\hat{\theta}_M){=}k] & \text{, if $\mathcal{A} = \texttt{modal}$}
\end{cases}, \ \forall j \in N_2, \forall k \in \mathcal{X}$ \Comment{Weights}

\If{$\mathcal{C} \in \{\texttt{BCH, ML}\}$}  \Comment{Bias correction methods}
    \State Define the r.v. $W_j$ with pmf $p(W_j{=}k) = w_{jk}, \ \forall j \in N_2, \forall k \in \mathcal{X}$\Comment{Predicted class membership}

    \State Set $D_{ck} = p(W{=}k|X{=}c; \hat{\theta}_M), \ \forall (c,k) \in \mathcal{X}^2$  \Comment{Estimated misclassification probabilities}
    
    \If{$\mathcal{C} = \texttt{BCH}$}
    \State $w_j = w_jD^{-1}, \ \forall j \in N_2$  \Comment{Bias-adjusted weights}
    
    \Else{}
    \State Set $w^*_j=w_jD^{T}, \ \forall j \in N_2$  \Comment{Class-conditional probability of predicted class membership}
    \EndIf
\EndIf

\vspace{-0.1cm}
\\\hrulefill
\State \underline{\textbf{STEP 3 : EM algorithm on the SM, based on imputed class weights}}
\State Set $t=0$

\Repeat
\vspace{-0.2cm}

    \State \hrulefill \text{ E Step }\hrulefill
    \If{$\mathcal{C} = \texttt{ML}$}
        \State $\tau^{(t)}_{jk} = w^*_{jk}p(X_j{=}k|z^o_j;z^p_j,\hat{\theta}_S^{(t)}), \ \forall j \in N_2, \forall k \in \mathcal{X}$ \Comment{Update responsibilities}
    \Else{}
         \State $\tau^{(t)}_{jk} = w_{jk} , \ \forall j \in N_2, \forall k \in \mathcal{X}$ \Comment{Imputed class weights used as responsibilities}
    \EndIf
    \State $\tau^{(t)}_{jk} = \frac{\tau^{(t)}_{jk}}{\sum_{c \in \mathcal{X}}\tau^{(t)}_{jc}} \ \forall j \in N_2, \forall k \in \mathcal{X}$ \Comment{Normalize responsibilities}

    \State \hrulefill \text{ M Step }\hrulefill

    \State $\hat{\theta}_{S}^{(t+1)} = \argmax\limits_{\theta_{S} \in \Theta_{S}}\sum\limits_{j\in N_2}\omega_j\sum\limits_{k\in\mathcal{X}} \tau^{(t)}_{jk}\log p(X_j{=}k, z^o_j;z^p_j,\theta_{S})$
    \Comment{Update $\hat{\theta}_{S}$}

    \State $t=t+1$
\Until{convergence}

\State Set $\hat{\theta}_S = \hat{\theta}_{S}^{(t)}$ \Comment{Estimated SM parameters}

\\\hrulefill
\State Return $(\hat{\theta}_M, \hat{\theta}_S)$ \Comment{Return estimated CM parameters}
\end{algorithmic}
\end{algorithm}

%% file: tables/distributions.tex
\renewcommand{\arraystretch}{2.5}
\begin{table}[t!]
\begin{center}
\begin{small}
\resizebox{\columnwidth}{!}{%
\begin{tabular}{llll}
\toprule
 \pkg{StepMix}              & \makecell[l]{\text{Observed} \\ \text{variable}} & \makecell[l]{\text{Parameters}\\ (for each $x \in \{1, ..., K\}$)}  & \text{PDF/PMF} \\
\midrule
 \code{binary}  & $z^o \in \{0, 1\}^D$ & $\pi_x \in [0, 1]^D$           & $p(z^o |x) = \prod_{d=1}^D z^o_d\pi_{x, d} + (1 - z^o_d)(1 - \pi_{x, d})$        \\
 \code{categorical} & \makecell[l]{$z^o \in \{0, \ldots, C\}^{D}$ }&  \makecell[l]{$\mathcal{P}_x \in [0, 1]^{D \times C}$ \\$ \sum_{c=1}^C \mathcal{P}_{x,d,c} = 1$}         &  $p(z^o |x) = \prod_{d=1}^D \mathcal{P}_{x,d,z^o_{d}} $       \\
 \code{gaussian\_unit} & $z^o \in \mathbb{R}^D$& $\mu_x \in \mathbb{R}^D$           & $p(z^o | x) = \mathcal{N}(z^o ; \mu_x, I)$        \\
 \code{gaussian\_spherical} & $z^o \in \mathbb{R}^D$& $\mu_x \in \mathbb{R}^D,\ \sigma^2_x \in \mathbb{R}$           &         $p(z^o | x) = \mathcal{N}(z^o ; \mu_x, \sigma^2_xI)$\\
 \code{gaussian\_diag}      & $z^o \in \mathbb{R}^D$& $\mu_x \in \mathbb{R}^D, \ \sigma^2_x \in \mathbb{R}^D$          &        $p(z^o | x) = \mathcal{N}(z^o ; \mu_x, diag(\sigma^2_x))$ \\
 \code{gaussian\_full}      & $z^o \in \mathbb{R}^D$& $\mu_x \in \mathbb{R}^D,\ \Sigma_x \in \mathbb{R}^{D \times D}$           & $p(z^o | x) = \mathcal{N}(z^o ; \mu_x, \Sigma_x)$    \\
 \code{covariate} & $z^p \in \mathbb{R}^D$&   $\beta_x \in \mathbb{R}^D,\ b_x \in \mathbb{R}$         & $p(x | z^p) = \frac{\exp(\beta_x^\top z^p + b_x)}{\sum_{k=1}^K \exp(\beta_k^\top z^p + b_k)}$       \\
\bottomrule
\end{tabular}}
\end{small}
\end{center}
\caption{Probability distributions supported by \pkg{StepMix}, and the associated strings to be passed as the \code{measurement} and/or \code{structural} arguments of the constructor (first column) for some observed variable of dimension $D$ and $K$ latent classes. \pkg{StepMix} can be used to fit a single set of observed variables (by only specifying \code{measurement}) and provides a syntax for combining continuous and categorical variables in a single model. For notational clarity, parameters for a single latent class are shown, and all $D$ categorical variables are integer-encoded. $\mathcal{N}(\cdot ; \mu, \Sigma)$ denotes the pdf of a Gaussian distribution parametrized by a mean $\mu$ and a covariance matrix $\Sigma$.}
\label{tab:StepMix_models}
\end{table}
\renewcommand{\arraystretch}{1}

%% file: tables/sim_response.tex
\begin{table}[t!]
\begin{center}
\begin{small}
\resizebox{\columnwidth}{!}{%
\begin{tabular}{lccccccccccr}
\toprule
\multirowcell{3}{Class\\ separation} & \multirowcell{3}{Sample\\ size} & \multicolumn{5}{c}{Mean bias} & \multicolumn{5}{c}{RMSE} \\
\cmidrule(lr){3-7} \cmidrule(lr){8-12}
 &  & \textit{1-step}&\textit{2-step}&\multicolumn{3}{c}{\textit{3-step}}&\textit{1-step}&\textit{2-step}&\multicolumn{3}{c}{\textit{3-step}} \\
\cmidrule(lr){5-7} \cmidrule(lr){10-12} & & & & Naive & BCH & ML & & & Naive & BCH & ML\\
\midrule
Low ($\gamma=0.7$)  & 500  &     .01 &  -.19 &          -.64 &        -.38 &       -.14 &   .16 &   .33 &           .66 &         .48 &        .31 \\
    & 1000 &                     .01 &  -.07 &          -.61 &        -.23 &       -.05 &   .11 &   .22 &           .63 &         .36 &        .22 \\
    & 2000 &                    -.00 &  -.02 &          -.57 &        -.11 &       -.01 &   .08 &   .17 &           .58 &         .24 &        .17 \\
Medium ($\gamma=0.8$) & 500  &   .00 &  -.03 &          -.31 &        -.04 &       -.02 &   .11 &   .15 &           .33 &         .18 &        .15 \\
                      & 1000 &  -.00 &  -.02 &          -.29 &        -.01 &       -.01 &   .08 &   .09 &           .30 &         .12 &        .09 \\
                      & 2000 &  -.00 &  -.01 &          -.28 &         .00 &       -.00 &   .06 &   .07 &           .29 &         .09 &        .07 \\
High ($\gamma=0.9$)   & 500  &  -.00 &  -.01 &          -.08 &         .00 &       -.00 &   .09 &   .09 &           .12 &         .09 &        .09 \\
                      & 1000 &  -.00 &  -.00 &          -.08 &         .00 &        .00 &   .07 &   .06 &           .10 &         .06 &        .06 \\
                      & 2000 &  -.00 &  -.00 &          -.08 &         .00 &       -.00 &   .05 &   .04 &           .09 &         .04 &        .04 \\
\bottomrule
\end{tabular}}
\end{small}
\end{center}
\vskip -0.1in
\caption{Simulation results for \pkg{StepMix} point estimates of one parameter in an outcome SM (with true value $\mu_2 = 1$). As described in the main text, three degrees of class separation in the MM (lower is harder) and three sample sizes are tested. Each configuration is simulated 500 times and the average parameter bias and RMSE are reported for one-, two- and three-step estimation, as well as the BCH and ML bias correction methods in the three-step case.}
\label{tab:sim}
\end{table}

%% file: tables/sim_response_bakk.tex
\begin{table}[t!]
\begin{center}
\begin{small}
\resizebox{\columnwidth}{!}{%
\begin{tabular}{lccccccccccr}
\toprule
\multirowcell{3}{Class\\ separation} & \multirowcell{3}{Sample\\ size} & \multicolumn{5}{c}{Mean bias} & \multicolumn{5}{c}{RMSE} \\
\cmidrule(lr){3-7} \cmidrule(lr){8-12}
 &  & \textit{1-step}&\textit{2-step}&\multicolumn{3}{c}{\textit{3-step}}&\textit{1-step}&\textit{2-step}&\multicolumn{3}{c}{\textit{3-step}} \\
\cmidrule(lr){5-7} \cmidrule(lr){10-12} & & & & Naive & BCH & ML & & & Naive & BCH & ML\\
\midrule
Low ($\gamma=0.7$)       & 500 &  .01& -.32& -.66& -.38& -.36& .19& .44& .70& .50& .47 \\
                         & 1000&  .01& -.16& -.60& -.22& -.20& .12& .30& .63& .37& .33 \\
                         & 2000&  .00& -.08& -.58& -.12& -.11& .08& .21& .59& .27& .23 \\
Medium ($\gamma=0.8$)    & 500&   .00& -.03& -.31& -.03& -.03& .11& .13& .33& .16& .13 \\
                         & 1000&  .00& -.01& -.29&  .01& -.01& .08& .09& .30& .12& .09 \\
                         & 2000&  .00&  .01& -.29&  .00&  .00& .05& .07& .29& .09& .07 \\
High ($\gamma=0.9$)      & 500&   .00&  .00& -.08&  .00&  .00& .08& .09& .12& .09& .09 \\
                         & 1000&  .01&  .01& -.07&  .01&  .01& .06& .06& .10& .07& .06 \\
                         & 2000&  .00&  .00& -.08&  .00&  .00& .04& .04& .09& .05& .04 \\
\bottomrule
\end{tabular}}
\end{small}
\end{center}
\vskip -0.1in
\caption{Simulation results for point estimates of one structural parameter of an outcome SM (with true value $\mu_2 = 1$) over 500 simulated data sets. Results are from  \citet{bakk2018two} and produced using \pkg{Latent GOLD}. Please refer to Table \ref{tab:sim} for simulation details. }
\label{tab:Bakk}
\end{table}

%% file: tables/sim_covariate.tex
\begin{table}[t!]
\begin{center}
\begin{small}
\resizebox{\columnwidth}{!}{%
\begin{tabular}{lccccccccccr}
\toprule
\multirowcell{3}{Class\\ separation} & \multirowcell{3}{Sample\\ size} & \multicolumn{5}{c}{Mean bias} & \multicolumn{5}{c}{RMSE} \\
\cmidrule(lr){3-7} \cmidrule(lr){8-12}
 &  & \textit{1-step}&\textit{2-step}&\multicolumn{3}{c}{\textit{3-step}}&\textit{1-step}&\textit{2-step}&\multicolumn{3}{c}{\textit{3-step}} \\
\cmidrule(lr){5-7} \cmidrule(lr){10-12} & & & & Naive & BCH & ML & & & Naive & BCH & ML\\
\midrule
Low ($\gamma=0.7$)    &  500 &   .03 &  -.27 &          -.63 &        -.29 &       -.31 &   .27 &   .41 &           .65 &         .57 &        .45 \\
                      & 1000 &   .01 &  -.15 &          -.61 &        -.19 &       -.18 &   .17 &   .29 &           .62 &         .42 &        .33 \\
                      & 2000 &  -.00 &  -.09 &          -.62 &        -.11 &       -.10 &   .12 &   .20 &           .62 &         .33 &        .23 \\
Medium ($\gamma=0.8$) &  500 &   .02 &  -.03 &          -.39 &        -.03 &       -.05 &   .16 &   .18 &           .40 &         .29 &        .21 \\
                      & 1000 &   .01 &  -.02 &          -.37 &         .01 &       -.02 &   .12 &   .13 &           .38 &         .21 &        .14 \\
                      & 2000 &  -.00 &  -.01 &          -.37 &         .01 &       -.01 &   .08 &   .09 &           .38 &         .14 &        .10 \\
High ($\gamma=0.9$)   & 500  &   .02 &   .01 &          -.12 &         .01 &        .01 &   .13 &   .13 &           .16 &         .15 &        .14 \\
                      & 1000 &   .01 &   .00 &          -.12 &         .01 &        .01 &   .09 &   .09 &           .15 &         .11 &        .10 \\
                      & 2000 &  -.00 &  -.00 &          -.13 &        -.00 &       -.00 &   .07 &   .07 &           .14 &         .07 &        .07 \\
\bottomrule
\end{tabular}}
\end{small}
\end{center}
\vskip -0.1in
\caption{Simulation results for \pkg{StepMix} point estimates of one parameter in a covariate SM (with true value $\beta_2 = 1$) over 500 simulated data sets. Simulation details are presented in the main text and in Table \ref{tab:sim}.}
\label{tab:sim_covariate}
\end{table}

%% file: tables/sim_covariate_bakk.tex
\begin{table}[t!]
\begin{center}
\begin{small}
\resizebox{\columnwidth}{!}{%
\begin{tabular}{lccccccccccr}
\toprule
\multirowcell{3}{Class\\ separation} & \multirowcell{3}{Sample\\ size} & \multicolumn{5}{c}{Mean bias} & \multicolumn{5}{c}{RMSE} \\
\cmidrule(lr){3-7} \cmidrule(lr){8-12}
 &  & \textit{1-step}&\textit{2-step}&\multicolumn{3}{c}{\textit{3-step}}&\textit{1-step}&\textit{2-step}&\multicolumn{3}{c}{\textit{3-step}} \\
\cmidrule(lr){5-7} \cmidrule(lr){10-12} & & & & Naive & BCH & ML & & & Naive & BCH & ML\\
\midrule
Low ($\gamma=0.7$)       & 500 &  .04& -.24& -.59& -.25& -.27& .25&   .38& .60& .48& .42 \\
                         & 1000&  .03& -.16& -.60& -.16& -.20& .19&   .32& .62& .45& .35 \\
                         & 2000&  .01& -.09& -.60& -.11& -.12& .12&   .22& .61& .36& .24 \\
Medium ($\gamma=0.8$)    & 500&   .01& -.05& -.40& -.03& -.06& .17&   .20& .41& .30& .22 \\
                         & 1000&  .01& -.01& -.37&  .02&   -.02& .11& .13& .38& .24& .14 \\
                         & 2000&  -.01&  -.01& -.37&  .01& -.01& .08& .09& .38& .15& .10 \\
High ($\gamma=0.9$)      & 500&   .02&  .01& -.11&  .02&  .01& .13&   .13& .17& .16& .14 \\
                         & 1000&  .01&  .01& -.12&  .01&  .01& .09&   .09& .15& .10& .10 \\
                         & 2000&  .00&  .00& -.12&  .01&  .00& .07&   .07& .13& .07& .07 \\
\bottomrule
\end{tabular}}
\end{small}
\end{center}
\vskip -0.1in
\caption{Simulation results for point estimates of one parameter in a covariate SM (with true value $\beta_2 = 1$) over 500 simulated data sets. Results are from  \citet{bakk2018two} and produced using \pkg{Latent GOLD}. Simulation details are presented in the main text and in Table \ref{tab:sim}.}
\label{tab:Bakk_covariate}
\end{table}

%% file: tables/sim_complete.tex
\begin{table}[t!]
\begin{center}
\begin{small}
\resizebox{\columnwidth}{!}{%
\begin{tabular}{lccccccccccr}
\toprule
\multirowcell{3}{Missing\\ values} & \multirowcell{3}{Sample\\ size} & \multicolumn{5}{c}{Mean bias} & \multicolumn{5}{c}{RMSE} \\
\cmidrule(lr){3-7} \cmidrule(lr){8-12}
 &  & \textit{1-step}&\textit{2-step}&\multicolumn{3}{c}{\textit{3-step}}&\textit{1-step}&\textit{2-step}&\multicolumn{3}{c}{\textit{3-step}} \\
\cmidrule(lr){5-7} \cmidrule(lr){10-12} & & & & Naive & BCH & ML & & & Naive & BCH & ML\\
\midrule
0\% &  500  &   .01 &  -.00 &          -.30 &        -.03 &        .00 &   .09 &   .11 &           .33 &         .18 &        .11 \\
     & 1000 &  -.00 &  -.00 &          -.29 &         .00 &       -.00 &   .06 &   .07 &           .30 &         .13 &        .08 \\
     & 2000 &  -.00 &  -.00 &          -.28 &         .01 &       -.00 &   .05 &   .05 &           .29 &         .09 &        .05 \\
25\% & 500  &   .00 &  -.04 &          -.45 &        -.13 &       -.00 &   .11 &   .17 &           .47 &         .28 &        .16 \\
     & 1000 &  -.00 &  -.01 &          -.43 &        -.08 &       -.00 &   .08 &   .10 &           .45 &         .21 &        .10 \\
     & 2000 &   .00 &   .00 &          -.41 &        -.03 &        .00 &   .05 &   .07 &           .42 &         .14 &        .07 \\
50\% & 500  &   .00 &  -.09 &          -.57 &        -.22 &       -.00 &   .14 &   .24 &           .60 &         .40 &        .20 \\
     & 1000 &   .00 &  -.04 &          -.57 &        -.16 &        .00 &   .10 &   .17 &           .59 &         .31 &        .15 \\
     & 2000 &   .00 &   .00 &          -.54 &        -.08 &        .01 &   .07 &   .12 &           .56 &         .23 &        .11 \\
\bottomrule

\end{tabular}}
\end{small}
\end{center}
\vskip -0.1in
\caption{Simulation results for \pkg{StepMix} point estimates of a SM's parameter in a model with six indicators, a covariate, and an outcome variable. We study the estimates of a single parameter in the outcome model (with true value $\mu_2 = 1$) over 500 simulated data sets. For this simulation, we fix the degree of class separation in the MM ($\gamma=0.8$) and instead vary the ratio of data missing completely at random in the indicators and the outcome. }
\label{tab:Bakk_complete}
\end{table}

%% file: tables/Real_data_table_MM.tex
\begin{table}[t!]
\begin{center}
\begin{small}
\begin{tabular}{lccc} 
\toprule
  Social status & Low & Middle & High \\ 
   \midrule
 Class size &   .70 & .23 & .07 \\ 
 Father’s job prestige &   &  &  \\ 
 \ \ Low &  .47 & .31 & .05 \\ 
 \ \ Medium &  .53 & .67 & .46 \\ 
 \ \ High &   .00 & .02 & .49 \\ 
 Mother’s education   &    &  &  \\ 
 \ \ Below high school  &   .82 & .14 & .15 \\  
 \ \ High school  &   .17 & .79 & .44 \\ 
 \ \ Junior college  &  .00 & .03 & .01 \\ 
 \ \ Bachelor &  .01 & .04 & .30 \\ 
 \ \ Graduate &  .00 & .01 & .10 \\ 
 Father’s education &    &  &  \\ 
 \ \ Below high school &   .95 & .06 & .00 \\  
 \ \ High school &   .05 & .89 & .11 \\ 
 \ \ Junior college &   .00 & .00 & .05 \\ 
 \ \ Bachelor &  .00 & .05 & .39 \\ 
 \ \ Graduate &  .00 & .00 & .44 \\  
\bottomrule
 
\end{tabular}
\end{small}
\end{center}
\caption{Estimated MM parameters (marginal distribution of social classes and class-conditional distribution of parents' education and job prestige).}
\label{tab:Real_data_MM}
\end{table}

%% file: tables/Real_data_table_SM.tex
\begin{table}[t!]
\begin{center}
\begin{small}
\begin{tabular}{lcccc} 
 \toprule
 Model &  & Low class income & Middle class income & High class income \\ 
  \midrule
 1-step &  & 20.32 (0.93) & 26.10 (3.02) & 67.90 (3.18)\\ 
 2-step &  & 25.25 (2.48) & 38.41 (6.14) & 50.66 (7.13) \\ 
 Naive 3-step &  & 27.44 (0.59) & 35.94 (1.28) & 43.67 (3.85) \\ 
 3-step BCH &  & 26.71 (0.74) & 36.81 (1.64) & 44.68 (4.26) \\ 
 3-step ML &  & 21.05 (1.52) & 44.73 (5.25) & 61.26 (11.77) \\ 
 
 \bottomrule
\end{tabular}
\end{small}
\end{center}
\caption{Estimated SM parameters (class-conditional average annual family income) and their standard error for each stepwise method.}
\label{tab:Real_data_SM}
\end{table}

%% file: tables/Real_data_table_SM_anova.tex
\begin{table}[t!]
\begin{center}
\begin{small}
\begin{tabular}{lcccc} 
\toprule
Model & Est. & SE & $Z$ & $P(>|z|)$ \\ 
  \midrule
1-step &  &  &  & \\ 
\ \ Middle class & 5.79 & 3.78 & 1.53 & $p=.126$ \\ 
\ \ High class & 47.58 & 3.49 & 13.63 & $p<.001$ \\ 
 
2-step &  &  &  & \\ 
\ \ Middle class & 13.16 & 8.38 & 1.57 & $p=.116$ \\ 
\ \ High class & 25.41 & 6.22 & 4.09 & $p<.001$ \\ 
 
Naive 3-step  &  &  &  & \\ 
\ \ Middle class & 8.50 & 1.41 & 6.02 & $p<.001$ \\ 
\ \ High class & 16.22 & 3.83 & 4.24 & $p<.001$ \\ 
 
3-step BCH   &  &  &  & \\ 
\ \ Middle class  & 10.11 & 1.93 & 5.24 & $p<.001$ \\ 
\ \ High class & 17.97 & 4.22 & 4.26 & $p<.001$ \\ 
3-step ML   &  &  &  & \\ 
\ \ Middle class & 23.68 & 6.27 & 3.77 & $p<.001$ \\ 
\ \ High class & 40.21 & 11.58 & 3.47 & $p<.001$ \\ 
  
\bottomrule
\end{tabular}
\end{small}
\end{center}
\caption{Family’s income differences between classes for each method.}
\label{tab:Real_data_SM_anova}
\end{table}

%% file: tables/comparison_packages.tex
\begin{table}[t!]
\begin{center}
\begin{small}
\resizebox{\columnwidth}{!}{%
\begin{tabular}{lcccccc}
\toprule
& \multicolumn{3}{c}{Latent class analysis}                                         & \multicolumn{3}{c}{Latent profile analysis}                                       \\
 \cmidrule(lr){2-4} \cmidrule(lr){5-7}      & Carcinoma          & Response simulation & Covariate simulation & Iris               & Diabetes            & Banknote           \\
Software        & \textit{MM only}            & \textit{1-step}              & \textit{1-step}               & \textit{MM only}            & \textit{1-step}              & \textit{1-step}             \\
\midrule
\pkg{StepMix} & -293.705  & -8600.793 & -5192.049   & -180.185  & -2407.146  & -771.669  \\
\pkg{poLCA}   & -293.705  & -                  & -5192.049   & -                 & -                  & -                 \\
\pkg{mclust}  & -                 & -                  & -                   & -180.186  & -                  & -                 \\
\pkg{MoEClust}  & -                 & -                  & -                   & -180.186  & -                  & -771.669               \\
\pkg{Mplus}   & -293.705    & -8599.609    & -5192.049     & -180.185     & -2407.146    & -771.668    \\
\bottomrule
\end{tabular}
}
\end{small}
\end{center}
\vskip -0.1in
\caption{Comparison of \pkg{StepMix} log-likelihoods with log-likelihoods from \pkg{poLCA}~\citep{linzer2011polca}, \pkg{mclust}~\citep{scrucca2016mclust}, \pkg{MoEClust}~\citep{MurphyandMurphy2020}, and \pkg{Mplus}~\citep{muthen2017mplus}. For latent class analysis (binary or categorical MM), we consider the Carcinoma dataset~\citep{agresti2002categorical} and the two simulation datasets from \citet{bakk2018two} discussed in Sections~\ref{Subsec:response_simulation} and \ref{Subsec:covariate_simulation}. For latent profile analysis (continuous MM), we benchmark the Iris~\citep{anderson1935irises, fisher1936use}, Diabetes~\citep{reaven1979attempt}, and Banknote~\citep{flury1988multivariate} datasets. 
We used three classes for each dataset except for Banknote, where we used two. Latent class models were computed with default MM parameters. In the case of latent profile models, we used diagonal covariance matrices for each class in the Diabetes and Banknote examples, which corresponds to \pkg{mclust}’s ``VVI'' model and \pkg{StepMix}’s \texttt{gaussian\_diag} model. For the Iris example, we used general covariance matrices for each class, which corresponds to \pkg{mclust}’s ``VVV'' model and \pkg{StepMix}’s \texttt{gaussian\_full} model. For \pkg{Mplus}, we manually defined the covariance structure to match other packages. The 1-step approach was used for all SMs to facilitate package comparison. Results were obtained using the default optimization parameters of each package, except in the Iris example, where 20 random initializations were used for \pkg{StepMix} and \pkg{Mplus}. The ``class'' variable was used as a categorical distal outcome in the Diabetes example, and ``status'' was used as a binary covariate in the Banknote example. All packages found solutions in under 1 second for all datasets. The ``-'' entries indicate models that are not supported and were thus not estimated by the corresponding package.}
\label{tab:comparison_packages}
\end{table}